\documentclass[12pt]{article}
\usepackage{hyperref}
\usepackage{amsmath, amsfonts, amssymb, mathrsfs}
\usepackage{dcolumn}
\usepackage{caption}
\usepackage{subcaption}
\usepackage{filemod}
\usepackage[ruled, vlined]{algorithm2e}
\usepackage{floatrow}
\usepackage{setspace}
\usepackage{verbatim}
\usepackage[Export]{adjustbox}
\usepackage{tabularx}
\usepackage{cellspace}
\usepackage{mathtools}
\usepackage{natbib}
\usepackage[title]{appendix}
\usepackage{bbm}
\usepackage{multirow}
\usepackage[dvipsnames]{xcolor}
\DeclareMathOperator*{\argmin}{arg\,min}
\hypersetup{
    colorlinks=true,
    linkcolor=blue, 
    urlcolor=black,
    citecolor=blue, 
    }

\title{\vspace{-0.3cm} Robust Wrapped Gaussian Process Inference for Noisy Angular Data}
\author{Andrew Cooper\thanks{Corresponding author: Virginia Tech, {\tt ahcooper@vt.edu}} 
	\and Justin Strait\thanks{Los Alamos National Laboratory} 
	\and Mary Frances Dorn\footnotemark[2]
    \and Robert B. Gramacy\thanks{Virginia Tech}
    \and Brendon Parsons\footnotemark[2]
    \and Alessandro Cattaneo\footnotemark[2]}
\date{\today}

\begin{document}

\vspace{-0.5cm}
\maketitle

\singlespacing

\vspace{-1.5cm}
\begin{abstract}
Angular data are commonly encountered in settings with a directional or orientational component.  Regressing an angular response on real-valued features requires intrinsically capturing the circular or spherical manifold the data lie on, or using an appropriate extrinsic transformation.  A popular example of the latter is the technique of distributional wrapping, in which functions are ``wrapped'' around the unit circle via a modulo-$2\pi$ transformation.  This approach enables flexible, non-linear models like Gaussian processes (GPs) to properly account for circular structure.  While straightforward in concept, the need to infer the latent unwrapped distribution along with its wrapping behavior makes inference difficult in noisy response settings, as misspecification of one can severely hinder estimation of the other.  However, applications such as radiowave analysis \citep{shangguan2015relative} and biomedical engineering \citep{kurz2015heart} encounter radial data where wrapping occurs in only one direction.  We therefore propose a novel wrapped GP (WGP) model formulation that recognizes monotonic wrapping behavior for more accurate inference in these situations.  This is achieved by estimating the locations where wrapping occurs and partitioning the input space accordingly.  We also specify a more robust Student's $t$ response likelihood, and take advantage of an elliptical slice sampling (ESS) algorithm for rejection-free sampling from the latent GP space.  We showcase our model's preferable performance on simulated examples compared to existing WGP methodologies.  We then apply our method to the problem of localizing radiofrequency identification (RFID) tags, in which we model the relationship between frequency and phase angle to infer how far away an RFID tag is from an antenna.
\end{abstract}


\singlespacing

\section{Introduction} \label{sec:intro}
Angular data are common in areas of research in which direction or orientation is measured.  Examples include the study of wind patterns in climatology \citep{carta2008joint, breckling2012analysis}, the movement of animals in zoology \citep{rivest2016general}, and diffusion and brain imaging in biomedicine \citep{mcgraw2006mises, ryali2013parcellation}.  Settings that exhibit a periodic component can also produce angular data.  Radiowave analysis often deals with angular data due to the cyclical nature of oscillating waves \citep{coleman2017analysis}.  The angle at which a radio antenna receives a radiowave signal, known commonly as the ``angle of arrival'', is one such example \citep{janaswamy2002angle}; another is the phase ``profile'', which has been studied for the purposes of localizing Radio Frequency Identification (RFID) tags \citep{nikitin2010phase, shangguan2015relative}.  Localizing has been of interest to RFID researchers for the purpose of quickly tracking and taking inventory of materials.

Communication between active antennas emitting radiowave signals and RFID tags which receive but do not emit signals (also known as ``passive'' tags) elicit a phase shift that depends on the frequency of the wave.  The difference in angles determined by the location in the wave's periodic cycle is also dependent on the distance between the tag and the antenna.  Inference of the relationship between phase angle and wave frequency may therefore provide a means of tag localization \citep{yang2021rfid}.  However, phase angle is known to be noisy due to multipath propagation in indoor environments.  This problem serves as the motivating example of our work: regressing a noisy angular response (phase angle) on a linear co-variate (wave frequency) for the purposes of inferring tag distance.

Common ``intrinsic'' approaches to angular regression (also known as ``circular'' regression) take advantage of parameterizations native to circular data, such as the von Mises distribution \citep{mardia1975statistics}.  Mean and scale parameters are regressed on linear inputs through appropriate link functions, akin to generalized modeling \citep{fisher1992regression, downs2002circular, gao2006application}.  While extensions in machine learning literature via artificial neural networks have been proposed \citep{laha2022}, these approaches can be limited by the unimodality of the von Mises curve.  Other intrinsic approaches such as geodesic regression and manifold learning, which model spatial dependencies between observations natively on the circle, have been proposed for circular regression \citep{fletcher2011geodesic}.  However, flexible non-linear models often require some form of approximation.  Predictions based on Gaussian fields can be applied to manifolds but require an appropriate tangent space approximation, for instance \citep{pigoli2016kriging}.  

We would like to flexibly model the phase angle-frequency relationship to provide localization predictions with appropriate uncertainty quantification (UQ).  One class of models known for having these qualities are Gaussian processes \citep[GPs;][]{williams2006gaussian}.  Popular in the spatial statistics and machine learning communities, GPs can capture non-linear behavior in response surfaces and provide intuitive UQ estimation. 
However, traditional GPs are only appropriate for continuous responses in Euclidean space.  Applying them to an angular response requires either projecting observations to the Euclidean plane \citep{mallasto2018wrapped}, or appropriately transforming the model itself.  A common example of the latter is known as distributional wrapping, an extrinsic modeling approach in which linear distributions are folded around the unit circle via a modulo-$2\pi$ transformation \citep{mardia2000directional, jammalamadaka2001topics, jammalamadaka2004new}.  Wrapping has been applied to a variety of common distributions, including the Gaussian \citep{bell2024review}.  Wrapped modeling is often chosen in favor of traditional intrinsic approaches that must specify an appropriate link function \citep{ferrari2009wrapping, ravindran2011bayesian}.  Similarly, wrapped Gaussian process (WGP) models have been proposed in spatial contexts for both univariate (circular) and bivariate (spherical) response surfaces \citep{jona2012spatial}.

While handy, the concept of wrapping produces inferential challenges that make estimation difficult.  WGPs assume responses are generated from an unobserved ``unwrapped'' GP that has been wrapped around the unit circle.  Direct GP inference requires knowledge of each observation's corresponding ``wrapping number'' in order to recover responses in the latent unwrapped space.  Inference therefore requires either marginalizing out all possible wrapping numbers (which entails infinite summation over the set of integers), or estimating the wrapping numbers directly.  Most existing implementations opt for the second approach, and often use truncation of the wrapping space to make estimation feasible \citep{jona2012spatial}.  However, since wrapping numbers scale with the training set, estimation can be challenging in large-data settings.  What's more, latent GP inference by itself is challenging when the number of training samples is large.  Methods like the Laplace approximation \citep{williams2006gaussian} and variational inference \citep{tran2015variational} can be effective for training, but tend to undercut UQ due to their approximation of the posterior form.  Fully Bayesian approaches can be achieved through Monte Carlo (MC) integration, but sampling from the posterior using algorithms like Metropolis-Hastings (MH) is computationally infeasible for large training sizes.  For this reason, current WGP approaches instead rely on estimated wrapping numbers to ``reconstruct'' the latent unwrapped values.
\begin{figure}[ht!]
    \centering
    \includegraphics[width=0.32\linewidth,trim=0 30 30 50, clip=TRUE]{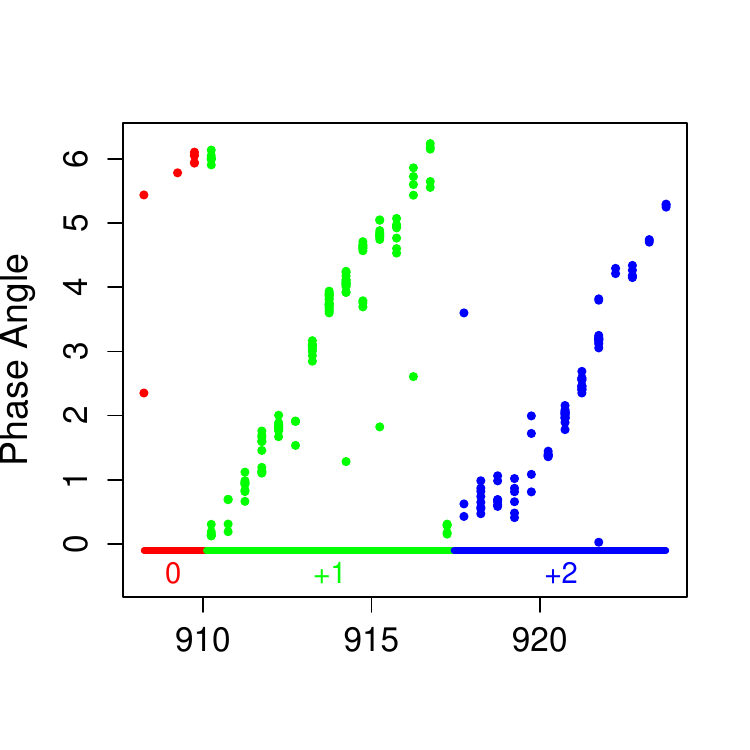}
    \includegraphics[width=0.32\linewidth,trim=0 30 30 50, clip=TRUE]{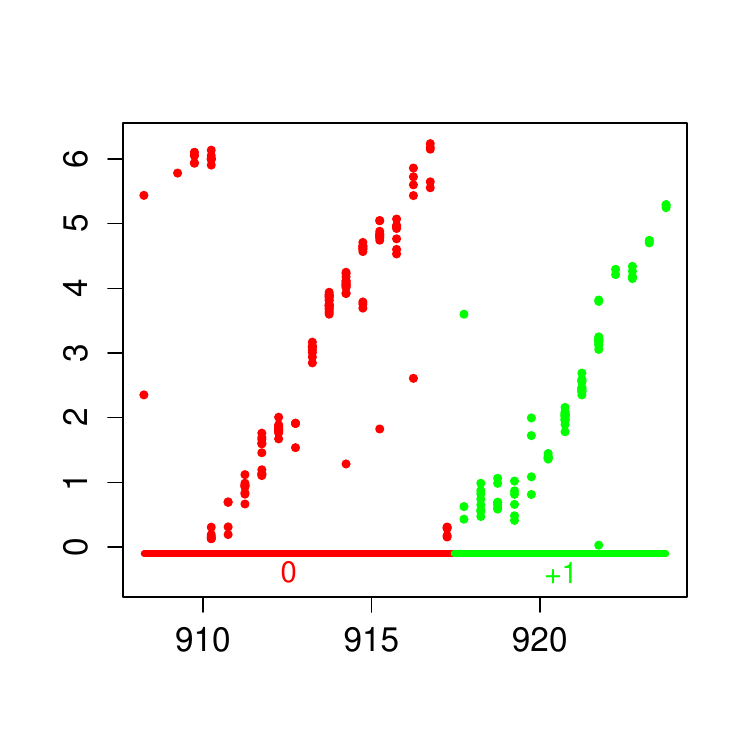}
    \includegraphics[width=0.32\linewidth,trim=0 30 30 50, clip=TRUE]{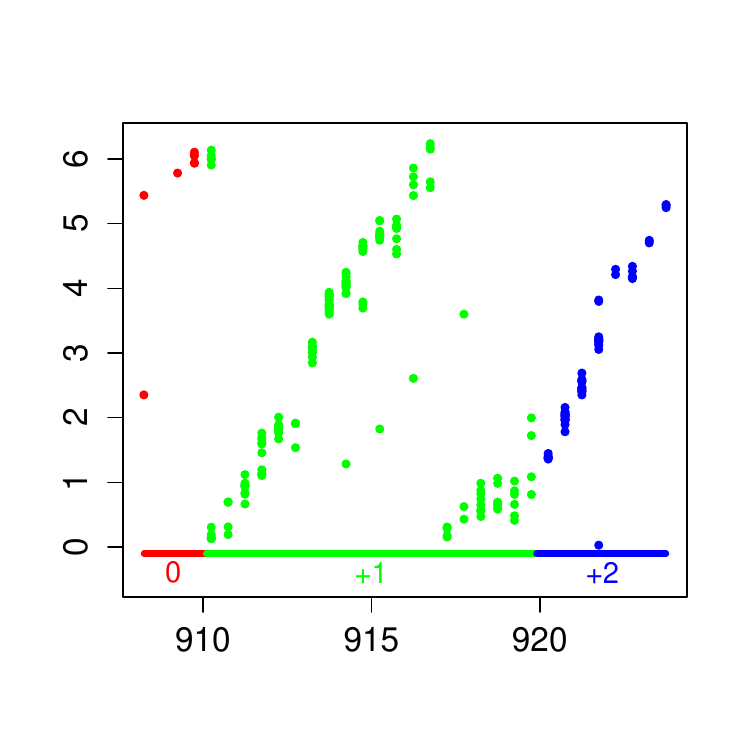}
    \includegraphics[width=0.32\linewidth,trim=0 10 30 40, clip=TRUE]{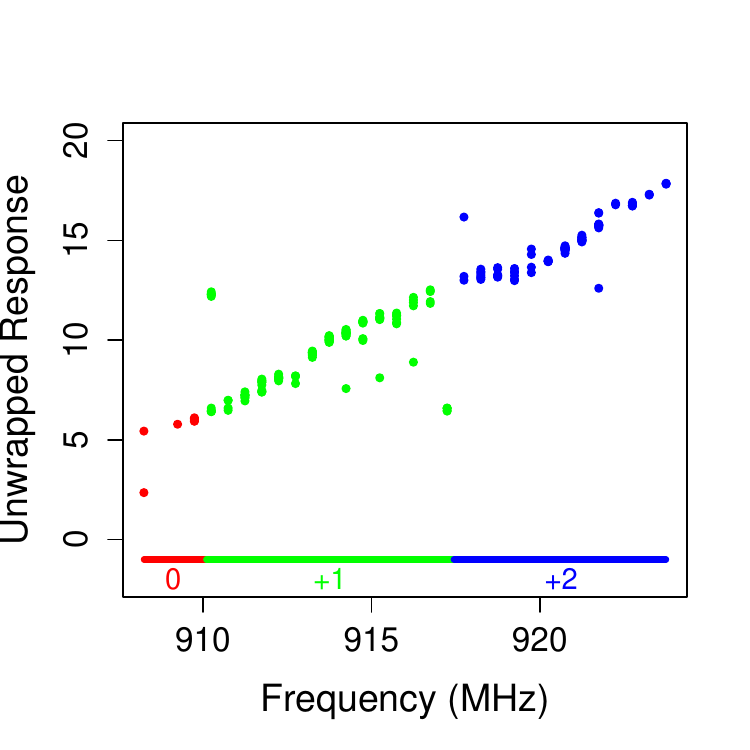}
    \includegraphics[width=0.32\linewidth,trim=0 10 30 40, clip=TRUE]{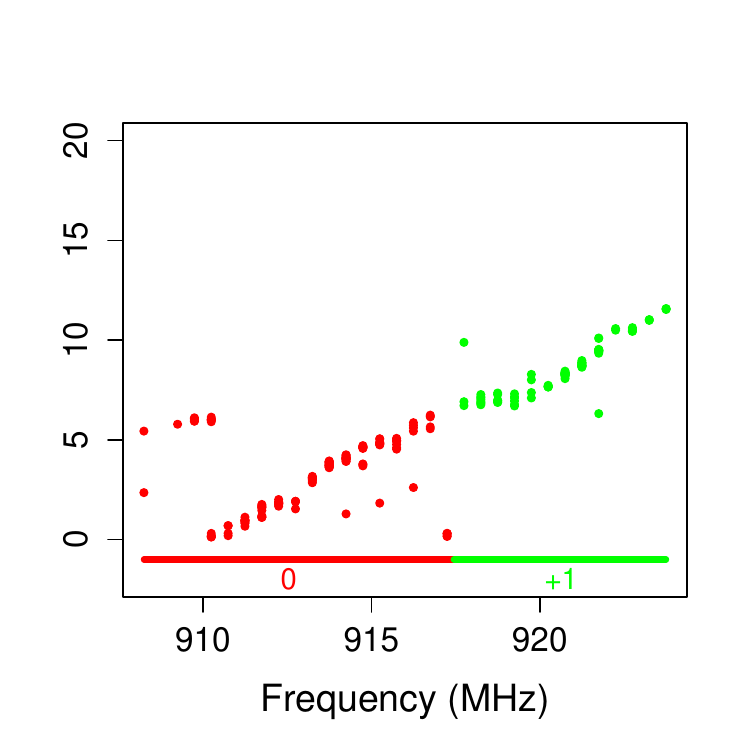}
    \includegraphics[width=0.32\linewidth,trim=0 10 30 40, clip=TRUE]{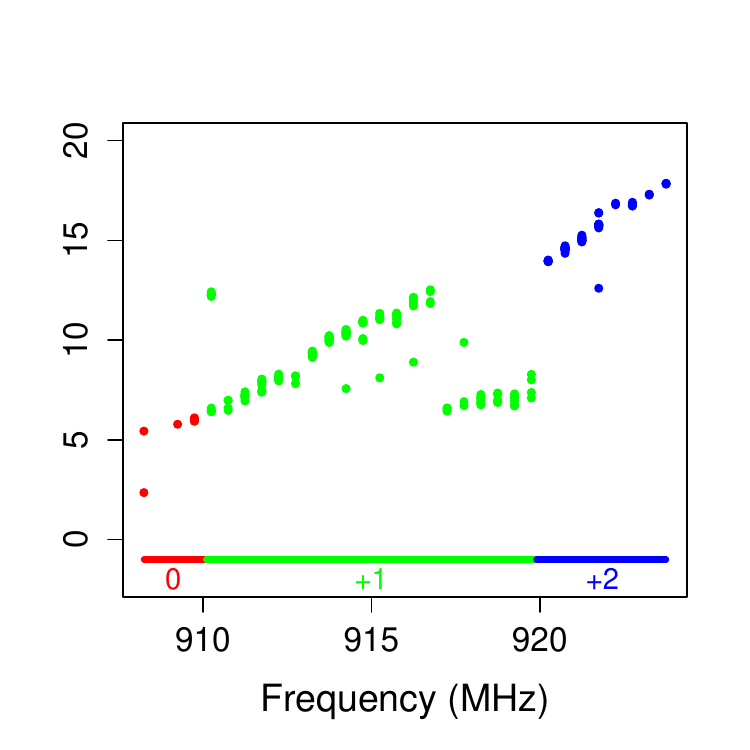}
    \caption{{\em Top:} Example phase-frequency data with three different estimates of where phase wraps around the unit circle for each observation. {\em Bottom:} The corresponding unwrapped data reconstructed in each scenario. The left version produces the ``smoothest'' looking unwrapped response, while the other two look more jagged due to poor estimation of the wrapping behavior.}
    \label{fig:rfid_ex}
\end{figure}

To illustrate how overreliance on wrapping number estimation can lead to inferential issues, consider an example relationship between phase and frequency shown in Figure \ref{fig:rfid_ex}.  Note the data plotted in the top three plots are identical, but each panel shows a different estimate of the wrapping numbers associated with the responses.  With these wrapping numbers, \cite{jona2012spatial} reconstruct the unwrapped response, which we plot in the bottom row.  As we see in the left column, proper estimation of the wrapping behavior leads to a relatively smooth unwrapped response surface, where GP estimation will be appropriate. Improper estimation, however, such as missing one of the wrapping locations (middle) or poorly estimating where the response wrapped around (right) leads to discontinuities in the unwrapped response, where GP inference will suffer.  Under this framework, WGP fitting is heavily reliant on proper inference of the wrapping behavior.  In addition to outliers observed in the wrapped data, improper estimation of the wrapping numbers creates structural outliers in the unwrapped space.  Models which assume independent Gaussian errors will likely not be able to accommodate these observations without specifying overly heavy tails.

We propose a novel WGP formulation that's more robust to both types of outliers by performing separate inference of both the wrapping numbers and the latent unwrapped output.  For the former, we take advantage of monotonic wrapping behavior in settings where the response exhibits a mean trend, such as what is seen in Figure \ref{fig:rfid_ex}.  By only estimating locations in the input space where the GP wraps around, rather than estimating wrapping numbers individually, we more efficiently explore the wrapping space without need for truncation.  For the latter, we use Elliptical Slice Sampling \citep[ESS;][]{murray2010elliptical} to directly sample from the unwrapped response.  ESS allows for feasible latent GP inference in large-data settings without the need for severe approximation.  Since ESS is agnostic to model likelihood, we specify a Student's $t$ distribution that's more flexible to outliers observed in Figure \ref{fig:rfid_ex}.  We combine these components into a single, fully Bayesian Markov-Chain Monte Carlo (MCMC) scheme, and empirically show that a ``decoupled'' estimation of the latent GP and corresponding wrapping numbers provides more robust inference in settings with a noisy response. 

The rest of this paper is organized as follows. In Section \ref{sec:background}, we review concepts foundational to WGP inference, including distributional wrapping and ordinary GP regression. In Section \ref{sec:wgp}, we describe our novel WGP formulation and detail both training and prediction for out-of-sample inputs.  In Section \ref{sec:sim}, we compare our method with other WGP implementations on simulated examples.  In Section \ref{sec:results}, we apply our WGP model to the unique problem of localizing RFID tags, and showcase its performance on experimental data collected in a lab environment.  
Section \ref{sec:discuss} concludes with some discussion.

\section{Background} \label{sec:background}
Here we introduce concepts relevant to our proposed WGP estimation procedure detailed in Section \ref{sec:wgp}.  We first overview distributional wrapping for angular modeling in Section \ref{ss:wrap}.  We then briefly review ordinary GP regression and its limitations when modeling angular response data in Section \ref{ss:gp_reg}.  Finally, we discuss WGPs for angular regression in Section \ref{ss:wrap_gp}, focusing on strategies for estimation proposed in the current literature.

\subsection{Wrapped Distributional Modeling} \label{ss:wrap}
Distributional wrapping is a common technique for modeling data that lie on a circular manifold \citep{mardia2000directional}.  One can take a random variable $Z$, whose distribution lies in Euclidean space, and ``wrap'' it around the unit circle: $Y = Z \bmod 2\pi$.  This transformation elicits a wrapped distribution $Y$ whose support lies in the interval $[0, 2\pi)$.  
\begin{figure}[ht!]
\centering
\includegraphics[scale=0.55, trim=0 40 0 20, clip=TRUE]{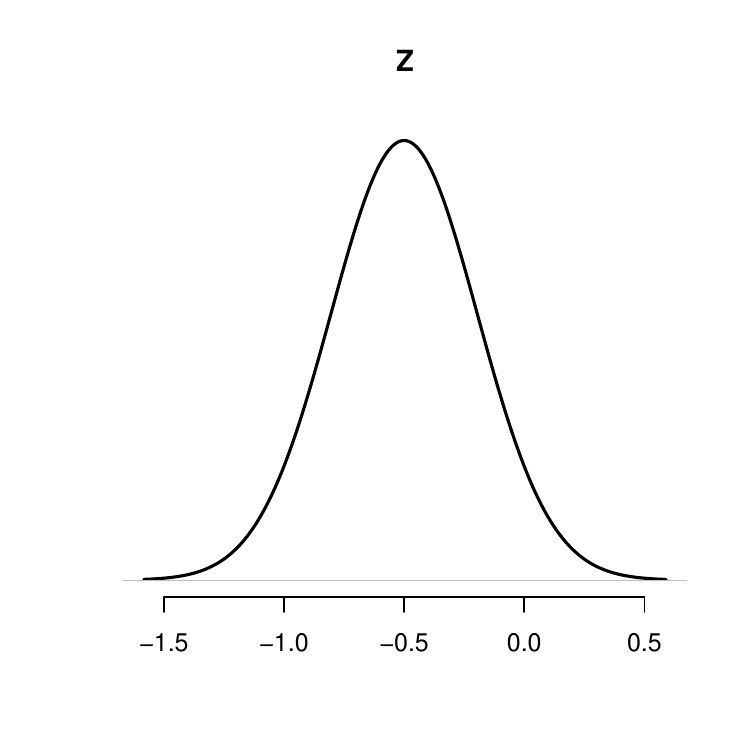}
\includegraphics[scale=0.55, trim=0 40 0 20, clip=TRUE]{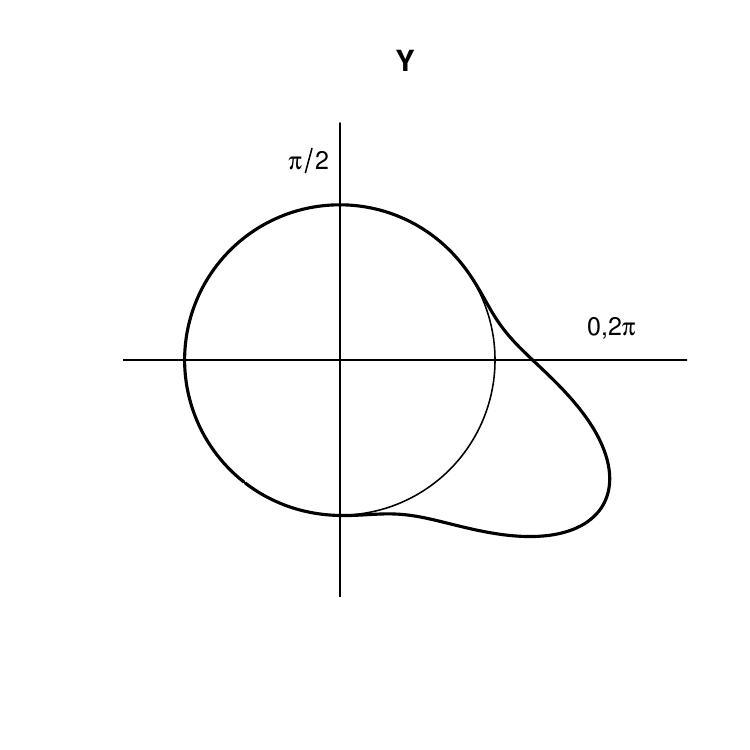}
\caption{{\em Left:} Gaussian distributed random variable $Z \sim \mathcal{N}(-0.5, 0.0025)$.  {\em Right:} The wrapped Gaussian random variable induced via the transformation $Y = Z \bmod 2\pi$.}
\label{fig:wrap_ex}
\end{figure}
For example, suppose $Z$ follows a Gaussian with mean $\mu$ and variance $\sigma^2$.  Then the transformed random variable $Y$ follows a Wrapped Gaussian (WG) with mean direction $\overset{\rightarrow}{\mu} = \mu - 2\pi K_\mu$ and concentration parameter $\rho = \exp\{-\sigma^2/2\}$.  The integer $K_\mu = \frac{1}{2\pi}\left(\mu - \overset{\rightarrow}{\mu}\right)$ denotes the number of times the mean for $Z$ wrapped around the unit circle.  For the illustrative example in Figure \ref{fig:wrap_ex}, $K_\mu = -1$ is negative since $\mu \in [-2\pi, 0)$.

Let $g$ represent the probability density function (PDF) for $Z$.  The PDF for $Y$ can be expressed as
\begin{equation}
p(y) = \sum_{k=-\infty}^{\infty}g\left(y + 2\pi k\right),
\label{eq:pdf}    
\end{equation}
where the doubly infinite sum stems from the many-to-one mapping of the modulo operation \citep{jammalamadaka2001topics}.  That is, there are an infinite number of unwrapped instances $z\in Z$ which correspond to the same wrapped instance $y = z \bmod 2\pi$.  Working backwards, one can recover $z$ via $z = y + 2\pi k$, where $k\in\mathbb{Z}$ corresponds to the wrapping number for $z$.  The wrapping number represents how many times, and in what direction, the unwrapped instance wrapped around the unit circle.  A negative wrapping number corresponds to clockwise wrapping around the circle, while positive corresponds to counter-clockwise.

Infinite summation over $k$ precludes exact evaluation of Eq.~(\ref{eq:pdf}), which has engendered the development of appropriate approximations.  The circular von Mises distribution has been shown to be a close approximation to the wrapped Gaussian \citep{jones2005family}.  Truncated series representations of the WG PDF have also been proposed \citep{kurz2014efficient}.  Other strategies involve estimating the latent wrapping number(s) directly. \cite{fisher1994time} proposed an Expectation-Maximization (EM) algorithm for inferring $k$, while \cite{coles1998inference} explored the space of wrapping numbers through individual MH acceptance-rejection steps with random walk proposals.  As we will discuss in Section \ref{ss:wrap_gp}, the issue of wrapping number inference is exacerbated in the case of angular regression, since it scales with the size of the training data.

\subsection{Gaussian Process Regression} \label{ss:gp_reg}
Gaussian processes (GPs) are flexible non-linear models commonly used in spatial statistics.  Consider $n$ real-valued inputs scaled to the unit interval $X = \{x_1,\dots,x_n\}\in [0,1]^n$, corresponding to $n$ outputs $Y = \{y_1,\dots,y_n\}\in \mathbb{R}^n$.  In this work we focus on cases with a one-dimensional input, although GPs can be used for inputs of arbitrary dimension.  An ordinary GP models the response a priori as instances of a smooth, continuous function $Y \sim \mathcal{N}_n\left(\mu(X), \tau^2\Sigma_\theta(X)\right)$, whose mean and covariance structure are dependent on $X$.  Often, $\mu(X)$ is set equal to zero if no inherent trend (such as linear) exists in $Y$.  The covariance matrix is commonly constructed using a kernel function, which is designed such that the pairwise covariances between elements of $Y$ are proportional to their (Euclidean) distances in $X$.  For instance, the squared-exponential kernel constructs the covariance matrix $\tau^2\Sigma_\theta(X)$ such that the $i$,$j^\text{th}$ entry of $\Sigma_\theta(X)$ is defined as
\begin{equation}
\Sigma_\theta(X)^{ij} = \Sigma_\theta(x_i, x_j) \equiv \exp\left\{-\dfrac{|x_i-x_j|^2}{\theta}\right\}.
\label{eq:sq_exp}    
\end{equation}
Although GPs are commonly known as ``non-parametric'' models, in addition to mean-specific parameters, kernel functions typically come with their own tuning hyperparameters.  In the case of Eq.~(\ref{eq:sq_exp}) we must specify a lengthscale $\theta > 0$, which determines the rate of correlation decay, and the outer scaling factor $\tau^2 > 0$, which determines the expected range of $Y$.
\begin{figure}[ht!]
\centering
\includegraphics[scale=0.6, trim=0 10 0 40, clip=TRUE]{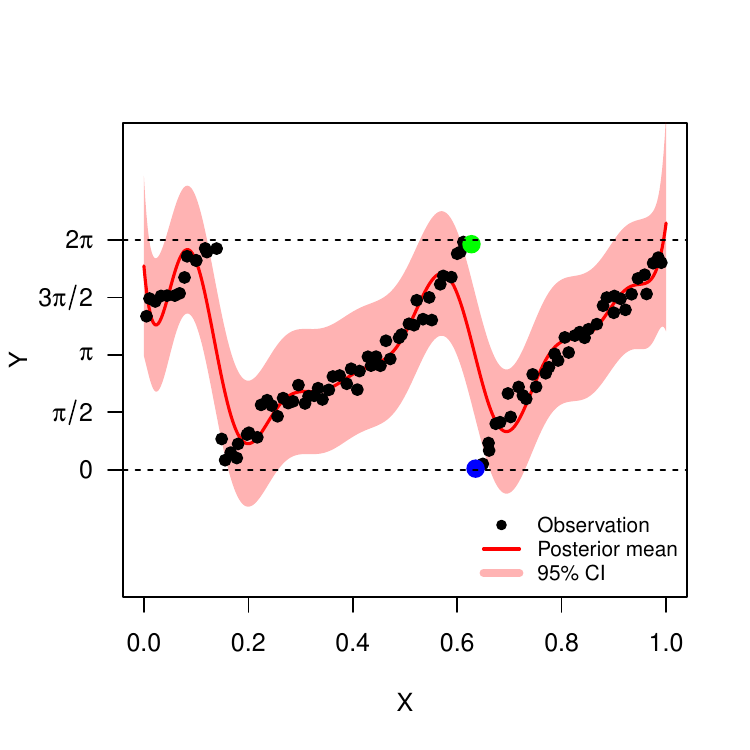}
\includegraphics[scale=0.58, trim=0 -20 0 10, clip=TRUE]{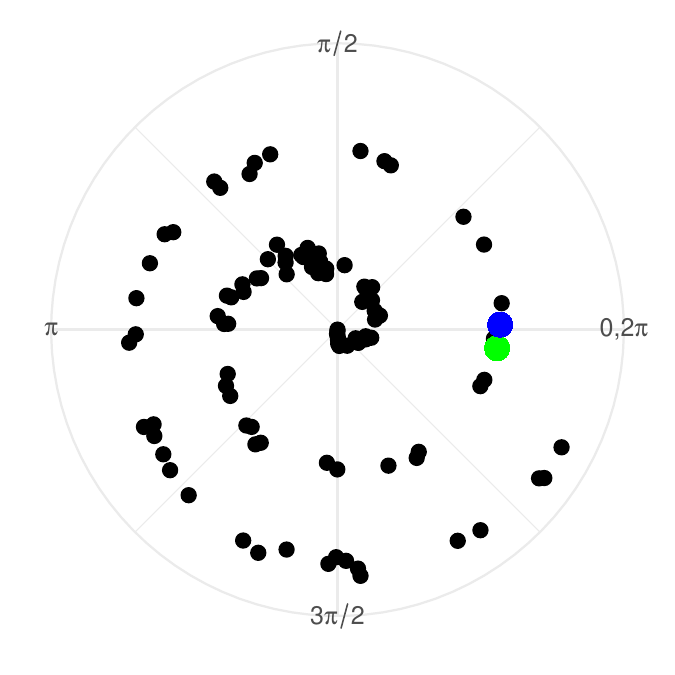}
\caption{{\em Left:} Simulated one-dimensional angular data from a Latin hypercube sampling (LHS) design of $n=50$ observations.  An ordinary GP was trained on this data and predicted on a fine grid of $n' = 400$ points, with a corresponding posterior mean and $95$\% credible interval (CI).  {\em Right:} The same observations plotted in polar coordinates.  Two points shown in green and blue are far away in the Euclidean plane but closer in the polar plane.}
\label{fig:gp_ex}
\end{figure}

Ordinary GPs assume $Y$ follows a multivariate normal (MVN) distribution with probability density function (PDF) $g$, where
\begin{equation}
g(Y) = \left(2\pi\tau^2\right)^{-\frac{n}{2}}|\Sigma_\theta(X)|^{-\frac{1}{2}}\exp\left\{-\dfrac{1}{2\tau^2}(Y-\mu(X))^\top\Sigma_\theta^{-1}(X)(Y-\mu(X))\right\}.
\label{eq:mvn}
\end{equation}
Gaussianity is inappropriate for radial data restricted to the interval $[0,2\pi)$; this can be seen in the simple one-dimensional example illustrated in the left panel of Figure \ref{fig:gp_ex}.  Not only do predictions fall outside the domain, the corresponding prediction uncertainty overestimates the variability in the response.  As mentioned earlier, the covariance structure depends on pairwise Euclidean distances between observations.  Two example observations shown in green and blue are close in $X$ but appear far when depicted in Euclidean space; however, note that on a circle manifold these observations are closer in $Y$.  This can be seen in the right panel, which plots the same data in polar coordinates via the transformation $\left(X\cos(Y), X\sin(Y)\right)$.  The overestimation of uncertainty stems from the ordinary GP failing to capture the circular structure of the response.

\subsection{Wrapped Gaussian Processes} \label{ss:wrap_gp}
To appropriately model responses that lie on a circular manifold, GPs can be ``wrapped'' around the unit circle \citep{jona2012spatial}.  Let $Z\sim\mathcal{N}_n\left(\mu(X),\tau^2\Sigma_\theta(X)\right)$ represent a latent GP as a function of $X$ with PDF $g$ defined in equation \ref{eq:mvn}.  Then $Y = Z \bmod 2\pi$ follows a wrapped Gaussian process (WGP) with PDF
\begin{equation}
p(Y) = \sum_{K\in\mathbb{Z}^n}g\left(Y + 2\pi K\right)\pi(K),
\label{eq:wgp}
\end{equation}
where $K = \{k_1,\dots,k_n\}\in \mathbb{Z}^n$.  We properly define its prior density $\pi$ in Section \ref{ss:wrap_num}.  The $n$-dimensional infinite sum contained in Eq.~(\ref{eq:wgp}) corresponds to summing over all potential wrapping numbers $k_i$ for each $y_i\in Y$.  Due to its intractability, most approaches to WGP inference use the joint PDF for $Y$ and $K$ instead, which follows the familiar MVN form of $g$.

Inference for $K$ is challenging due to its potential high dimension and fact that each value can take on any integer, positive or negative.  \cite{jona2012spatial} propose a ``reasonable'' truncation of the set of possible values for $K$ determined by an empirical estimate of the variance of the response.  Wrapping numbers are then sampled from their discrete finite posterior distribution iteratively in a MCMC scheme.  However, their method of truncation neglects the potential impact of a mean structure on the wrapping numbers.  In the case of a linear mean $\mu(X) = \alpha + \beta X$, the range of each $k_i$ depends on not only the variance of $Y$ but the slope $\beta$.  
We can see this visually in the left panel of Figure \ref{fig:gp_ex}.  A positive slope causes the response to wrap around the unit circle multiple times in the same direction; increasing the slope would likely lead to larger wrapping numbers, and vice versa.  \cite{coles1998inference} proposed MH sampling with a random walk proposal in the context of wrapped Gaussian modeling.  While this precludes the need for truncation, it requires $n$ separate MH steps corresponding to each entry in $K$;  this can make training prohibitively expensive in large-data settings.

As \cite{jona2012spatial} note, estimation of $K$ in WGP modeling is purely in service of inferring the latent, unwrapped GP $Z$.  That is, conditional on $K$, $Z$ can be constructed via $Y + 2\pi K$.  The authors take advantage of this to estimate mean and covariance-specific parameters.  However, what if the relationship between $(Y,K)$ and $Z$ is not noiseless?  Consider an additive noise term $\epsilon$ to $Y = (Z - 2\pi K + \epsilon)\text{mod }2\pi$.  The presence of noise ``post-wrapping'' means $Z$ cannot be reconstructed directly from $Y$ and $K$.  
\begin{figure}[ht!]
\centering
\includegraphics[width=0.325\textwidth, trim=0 15 0 55, clip=TRUE]{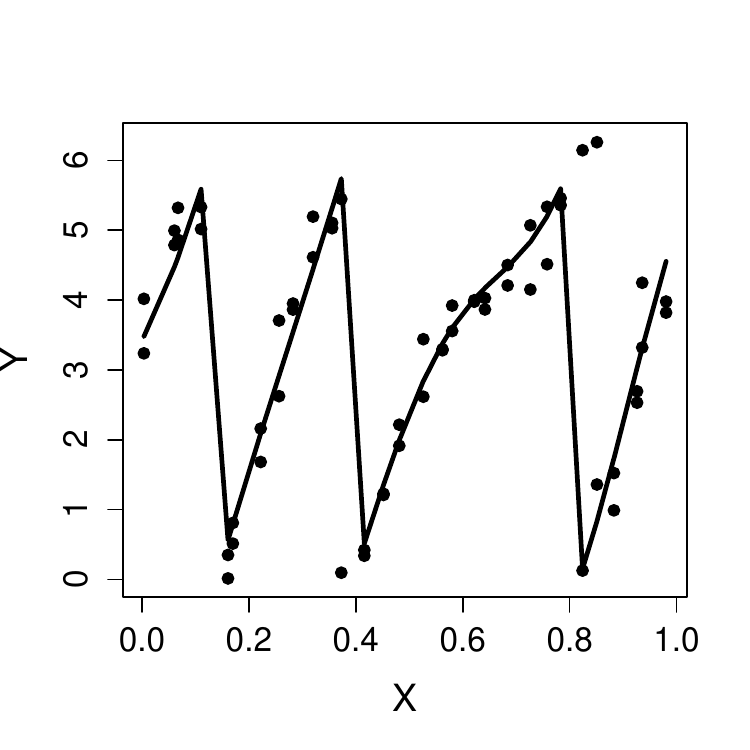}
\includegraphics[width=0.325\textwidth, trim=0 15 0 55, clip=TRUE]{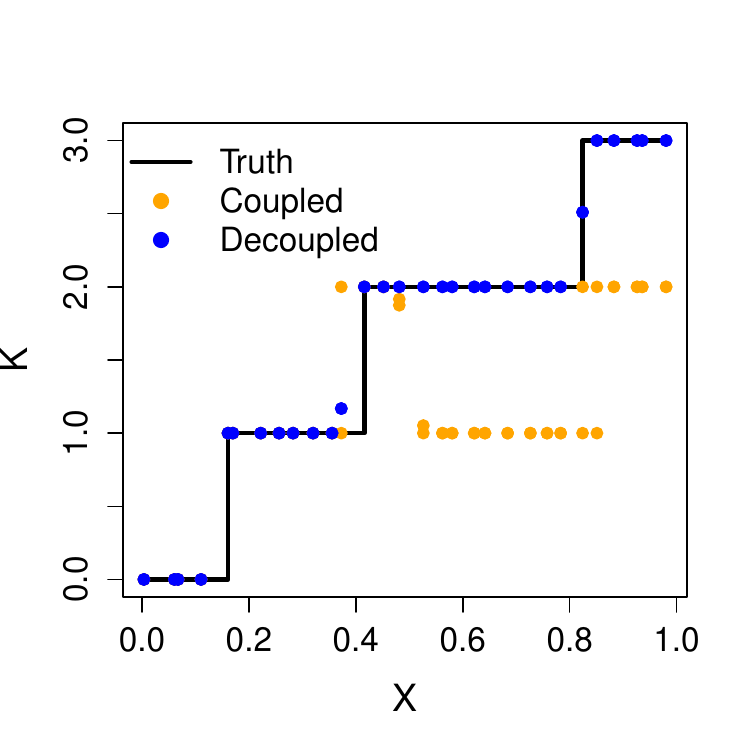}
\includegraphics[width=0.325\textwidth, trim=0 15 0 55, clip=TRUE]{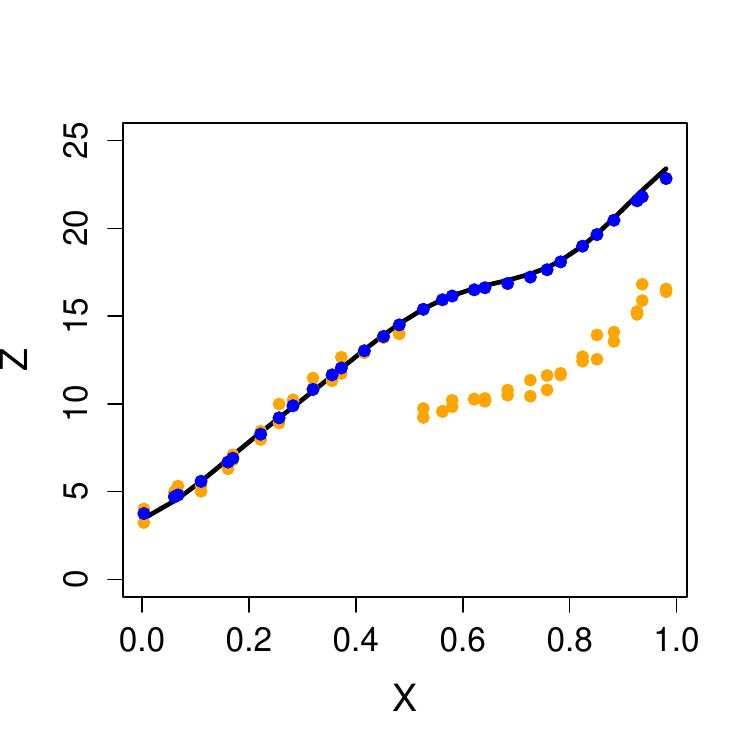}
\caption{{\em Left:} Example instance from a WGP (black), with $n=50$ observations with added independent Gaussian noise $\epsilon$.  {\em Middle:} The true known wrapping numbers $K$ (black), along with posterior estimation from a ``coupled'' approach that induces $Z$ from $K$ in orange \citep{jona2012spatial}, and a decoupled approach that estimates $K$ and $Z$ separately in blue (ours). {\em Right:} Both methods' reconstruction of the latent unwrapped GP compared to the truth.}
\label{fig:noise}
\end{figure}

As an illustrative example, consider radial responses generated from a WGP instance shown by the black line in the left panel of Figure \ref{fig:noise}.  The responses themselves contain independent additive Gaussian noise with variance $\sigma^2 = 0.2$.  Responses close to the $0$ and $2\pi$ boundaries may wrap around the unit circle due to this noise, causing them to appear to ``jump'' to either the top or the bottom of the plot.  Methods that assume a one-to-one relationship between $K$ and $Z$ can struggle to identify where the true latent GP wraps around the unit circle.  This can be seen in the middle panel, which compares the true known wrapping behavior with posterior mean estimates for $K$ from two different WGP methods.  First, a ``coupled'' estimation approach proposed by \cite{jona2012spatial} individually samples wrapping numbers from their discrete posterior distributions, and are evaluated based on the MVN likelihood for $Y + 2\pi K$.  However, when noise is present in $Y$, the unwrapped GP will appear noisier than in reality, which can negatively impact inference for $K$.  Poor estimation of $K$ consequently leads to improper reconstruction of $Z$, shown in the right panel.  Alternatively, a decoupled approach in which $K$ and $Z$ are estimated separately can identify the noiseless unwrapped process more accurately.  That is, allowing $K$ and $Z$ to vary separately, rather than inducing one quantity from the other, enables more robust estimation of WGPs.

\section{Wrapped Gaussian Process Inference} \label{sec:wgp}
In this section we propose our novel formulation for a one-dimensional WGP model for radial outputs that exhibit monotonic wrapping behavior in $X$:
\begin{align}
    Y &= (Z - 2\pi K + \epsilon) \bmod 2\pi &      \label{eq:wrap_model}
     \epsilon_1,\dots,\epsilon_n &\stackrel{\text{iid}}\sim t_\nu(0;\sigma^2) \\
    Z &\sim \mathcal{N}_n\left(\alpha + \beta X, \tau^2\Sigma_\theta(X) \right). \nonumber
\end{align}
We assume $Y$ is equal to a latent GP output $Z$ minus the corresponding wrapping numbers $2\pi K$, plus independent, identically distributed error term $\epsilon$.  This formulation allows for two important innovations. The first involves separate variables $K$ and $Z$, for which we specify independent priors.  The second is a specification of non-Gaussian distributed errors for $Y$; we prefer a central $t$-distribution with $\nu$ degrees of freedom and a variance inflation factor $\sigma^2$.  This elicits a likelihood
\begin{equation}
\mathcal{L}(Z,K\mid Y) = \prod_{i=1}^n \dfrac{\Gamma\left(\dfrac{\nu+1}{2}\right)}{\sqrt{\pi\nu\sigma^2}\Gamma\left(\dfrac{\nu}{2}\right)}\left(1+\dfrac{(y_i - (z_i-2\pi k_i))^2}{\nu\sigma^2}\right)^{-\frac{\nu+1}{2}},
\label{eq:like}
\end{equation}
where $\Gamma(x) = \int_0^\infty t^{x-1}e^{-x}dt$.  This choice of likelihood provides more robustness to outliers that violate the assumption of independent normally distributed errors \citep{lange1989robust}, as well as those caused by misspecifying the wrapping behavior.  The remaining subsections outline our fully Bayesian approach to posterior integration via for wrapping numbers $K\mid Z$ in Section \ref{ss:wrap_num}, latent and $Z \mid K$ via ESS in Section \ref{ss:ess}.  Section \ref{ss:wrap_fit} combines these two steps with
inference for other hyperparameters, and completes the description with a scheme for prediction at new inputs. 

\subsection{Monotonic Wrapping Number Estimation} \label{ss:wrap_num}
The main difficulty with wrapped GP inference is estimating $K$, as it requires exploring an $n$-dimensional space of both positive and negative integers.  However, rather than truncate the domain of $K$ \citep{jona2012spatial}, we propose a more efficient sampling procedure that takes advantage of known structure.
\begin{figure}[ht!]
\centering
\includegraphics[width=0.32\textwidth, trim=0 15 0 55, clip=TRUE]{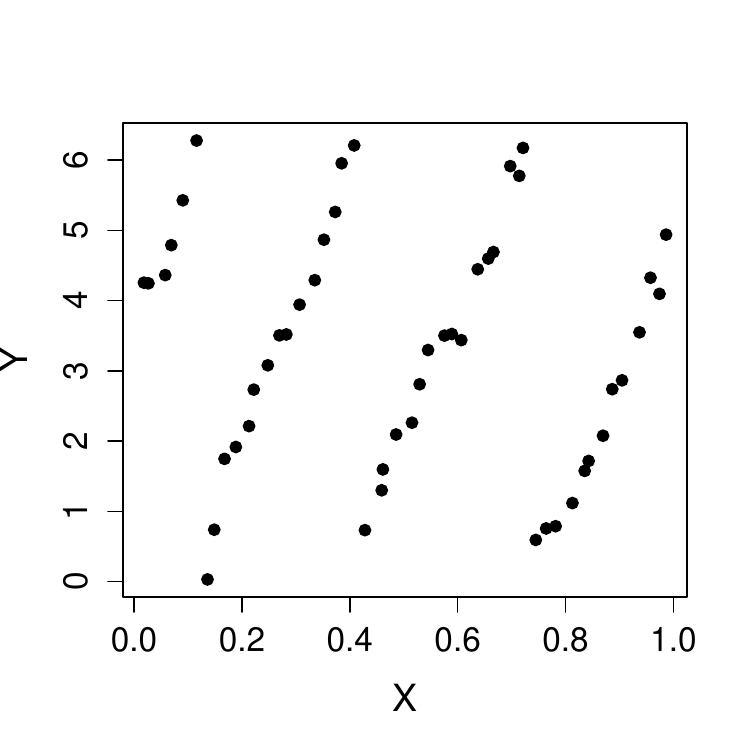}
\includegraphics[width=0.32\textwidth, trim=0 15 0 55, clip=TRUE]{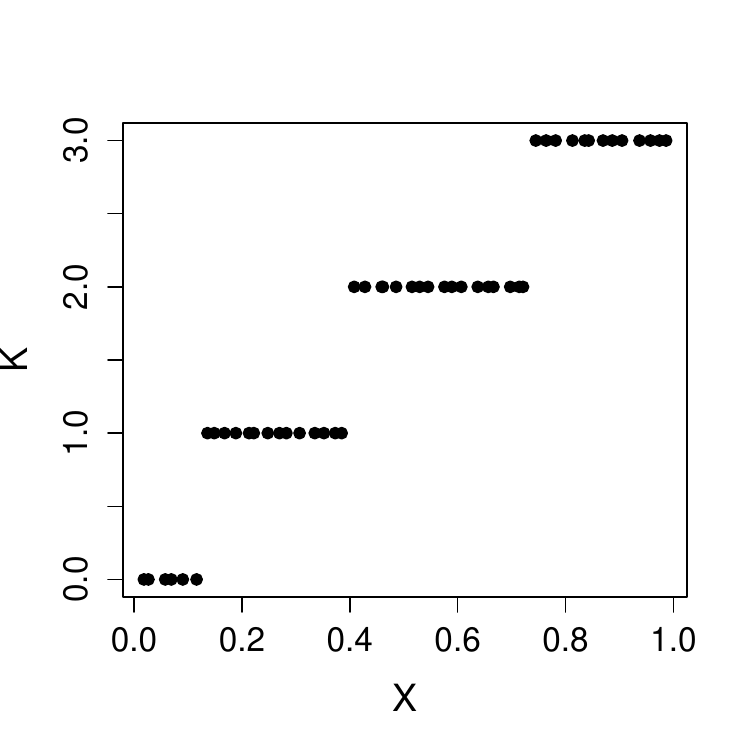}
\includegraphics[width=0.32\textwidth, trim=0 15 0 55, clip=TRUE]{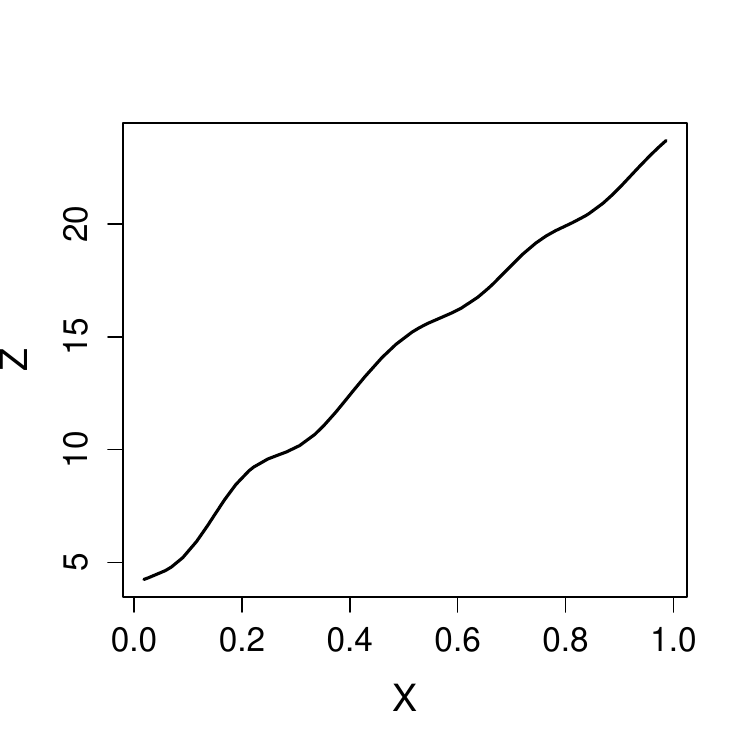}
\caption{{\em Left:} Observations generated from a WGP instance, generated from an LHS design of size $n=50$.  {\em Middle:} The wrapping numbers for each observation.  {\em Right:} The unwrapped GP output $Z$.}
\label{fig:k_ex}
\end{figure}
Consider the motivating simulated example shown in Figure \ref{fig:k_ex}.  Observations plotted in the left panel were generated from a linear mean WGP instance; the corresponding wrapping numbers are shown in the middle panel, while the latent unwrapped GP is shown on the right.  Data with a linear mean structure (left panel) typically ensures the response wraps in one direction (in this case, counter-clockwise), which results in step-wise behavior in the wrapping numbers (middle panel).  Assuming this structure a priori --- that is, $Y$ has a positive relationship with $X$ such that wrapping numbers are non-decreasing with $X$ (or conversely, a negative relationship with $X$ produces non-increasing behavior in $K$) --- we can establish important monotonic restrictions for $K$.  Specifically, in the case of counter-clockwise wrapping, if $x_i \leq x_j$ then $k_i \leq k_j$ for all $i\neq j$.  We also assume if $x_i = x_j$ then $k_i = k_j$; that is, replicates for the same input should have the same corresponding wrapping number, and differences in their response can be attributed to ``post-wrapping'' noise.  

This prior restriction drastically reduces the space of wrapping numbers that need to be considered.  Rather than entertain potentially infinitely supported sets of values for each element in $K$, we need only consider the values of $X$ where $Y$ wraps around the unit circle.  In the case of Figure \ref{fig:k_ex}, these ``wrapping locations'' occur around $x=0.15$, $0.4$ and $0.75$.  Conditional on those locations, wrapping numbers for observations contained in these ranges ($X\in (0,0.15)$, $(0.15,0.4)$, $(0.4,0.75)$, and $(0.75, 1)$) are constrained to one possible value due to the informative prior.
\begin{figure}[ht!]
\centering
\includegraphics[scale=0.55, trim=0 10 0 40, clip=TRUE]{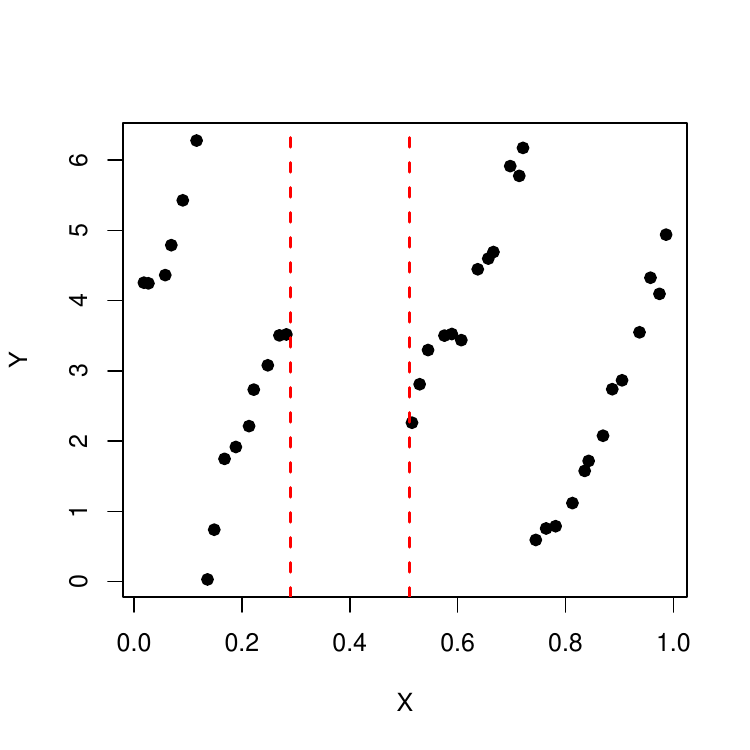}
\includegraphics[scale=0.55, trim=0 10 0 40, clip=TRUE]{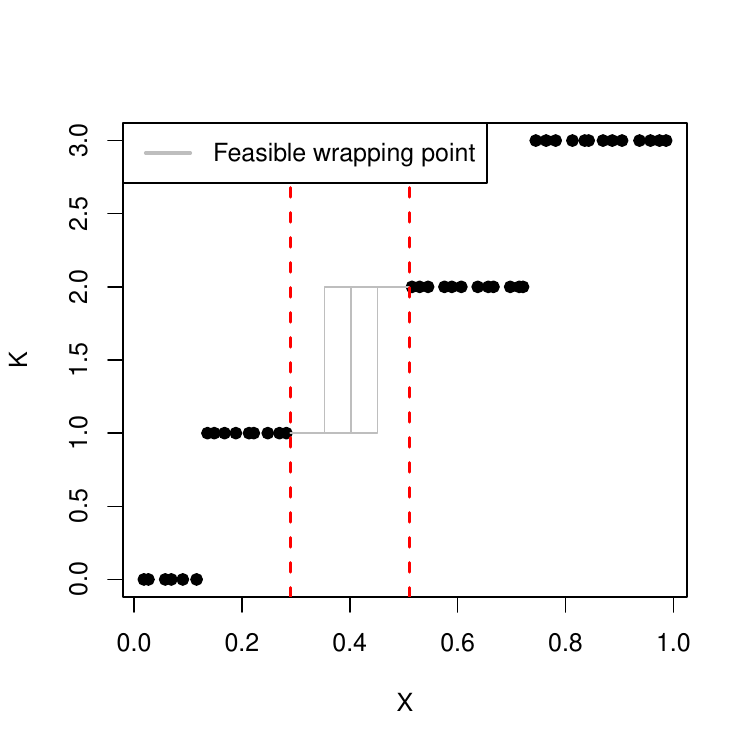}
\caption{No data is observed at inputs contained by the dashed red lines ({\em left}).  This leads to ambiguity in $K$, in which wrapping could occur anywhere in this region ({\em right}).}
\label{fig:missing}
\end{figure}

However, specifying $K$ solely for observed values of $X$ can lead to issues when there are gaps in the training data.  For instance, consider the same illustrative example shown in the left panel of Figure \ref{fig:missing}, where data is removed for $0.3\leq x \leq 0.5$.  In this setting, we can reason wrapping did occur due to the increase in $k$ from zero to one, but we cannot determine where wrapping occurs in this region, which consequently leads to ambiguity in the wrapping number space, as seen in the right panel.  That is, any proposed wrapping location between $0.3$ and $0.5$ has an equal likelihood of producing $Y$.  This non-identifiability issue for $K$ is not an issue for training purposes, but is important when it comes to performing proper UQ for out-of-sample prediction.

To this end, we propose a tree-based approach to $K$ inference that partitions the input space based on where distributional wrapping occurs.  Proper identification of the wrapping locations induces an appropriate value for $K$, which can be used to evaluate the likelihood given in Eq.~(\ref{eq:like}).  Borrowing some notation from \cite{chipman2010bart}, we define a function $k(x;W) \equiv \sum_{j=1}^m \mathbbm{1}(x \geq w_j) + k_{\min}$ which maps an input $x\in X$ to an integer $k\in\{k_{\min},\dots,k_{\min}+m\}$ based on $m$ wrapping locations $W=\{w_1,\dots,w_m\}$, where $w_i < w_j$ for $i<j$.  Wrapping numbers can then be induced by evaluating $k_i = k(x_i; W)$ for $i=1,\dots,n$.  We can represent this function equivalently as a binary decision tree, where wrapping locations serve as splitting criteria, and wrapping numbers are contained in the leaf nodes.  Proper specification of $k_{\min}$, the smallest possible value for $k_i$, is discussed in Appendix \ref{app:kmin}.

\begin{figure}[ht!]
\centering
\includegraphics[width=0.32\textwidth, trim=0 10 0 20, clip=TRUE]{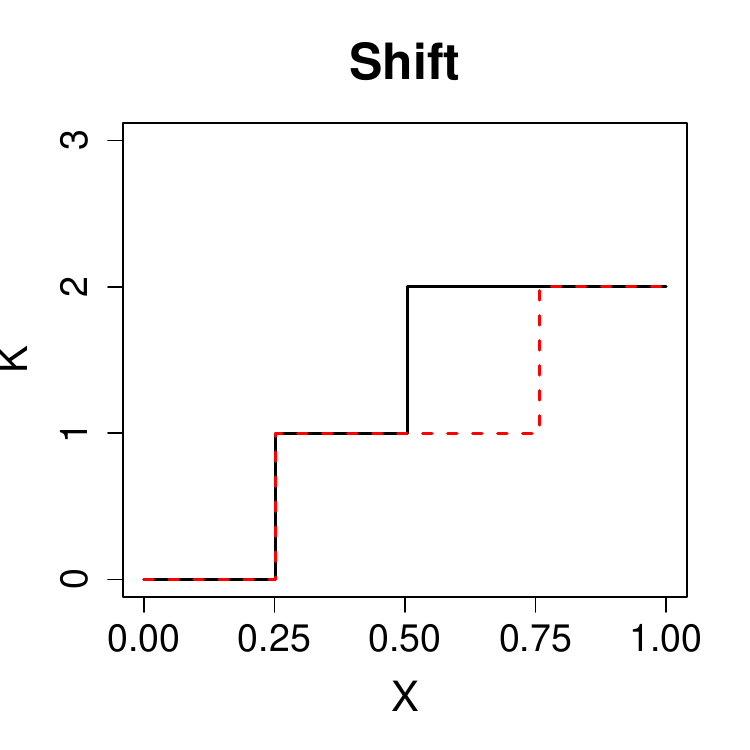}
\includegraphics[width=0.32\textwidth, trim=0 10 0 20, clip=TRUE]{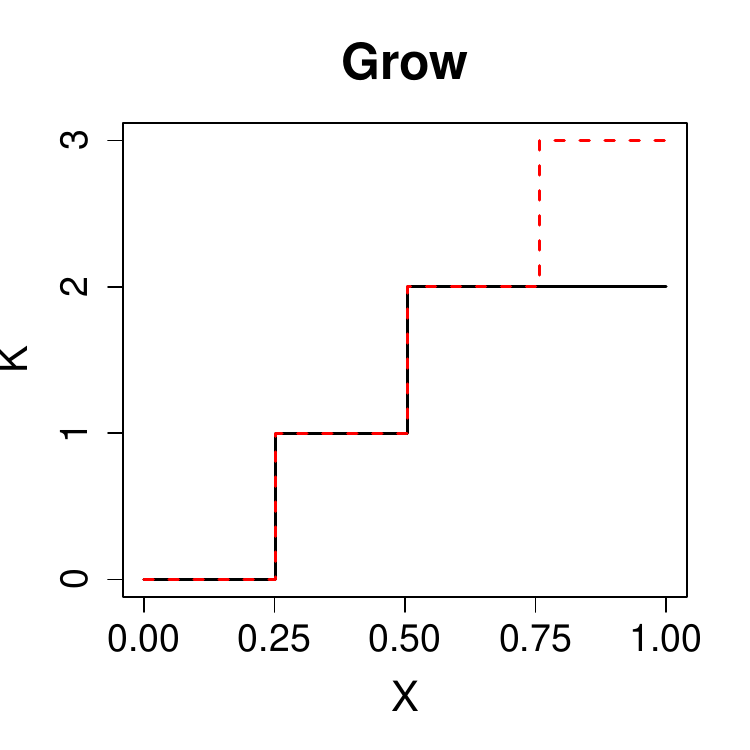}
\includegraphics[width=0.32\textwidth, trim=0 10 0 20, clip=TRUE]{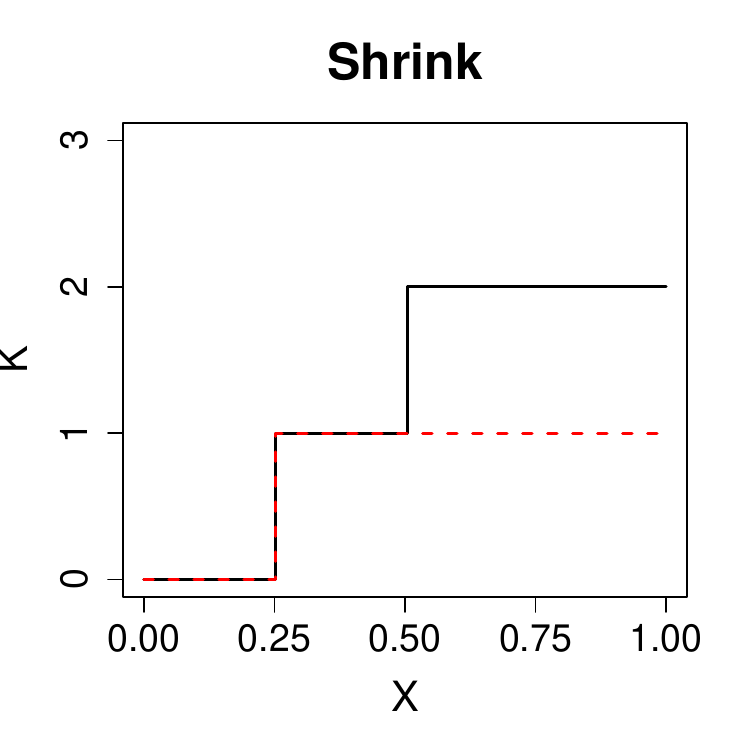}
\begin{minipage}{0.3\textwidth}
\hspace{0.75cm}
\includegraphics[scale=0.3, trim=0 -50 0 0, clip=TRUE]{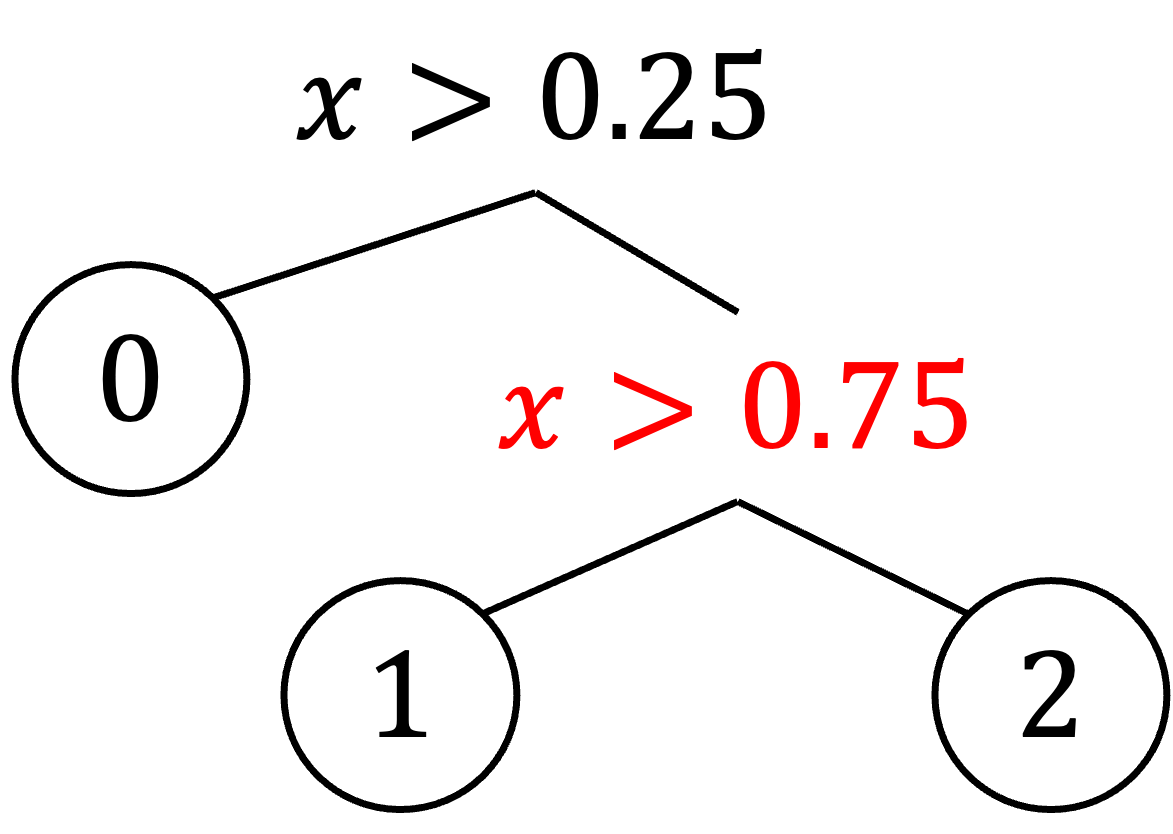}
\end{minipage}
\begin{minipage}{0.3\textwidth}
\hspace{1cm}
\includegraphics[scale=0.3, trim=0 0 0 -20, clip=TRUE]{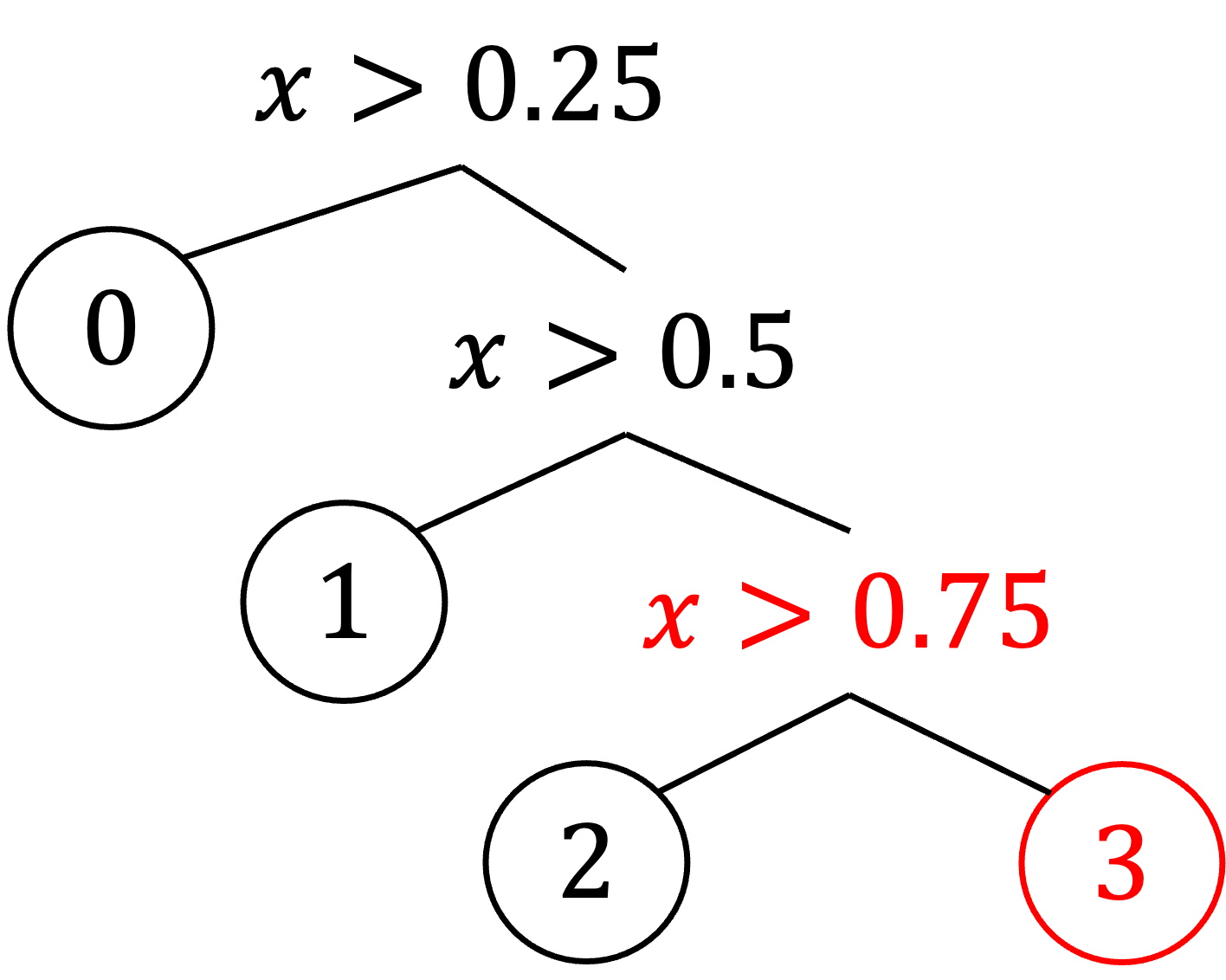}
\end{minipage}
\begin{minipage}{0.3\textwidth}
\vspace{-1cm}
\hspace{1.25cm}
\includegraphics[scale=0.3, trim=0 -20 0 0, clip=TRUE]{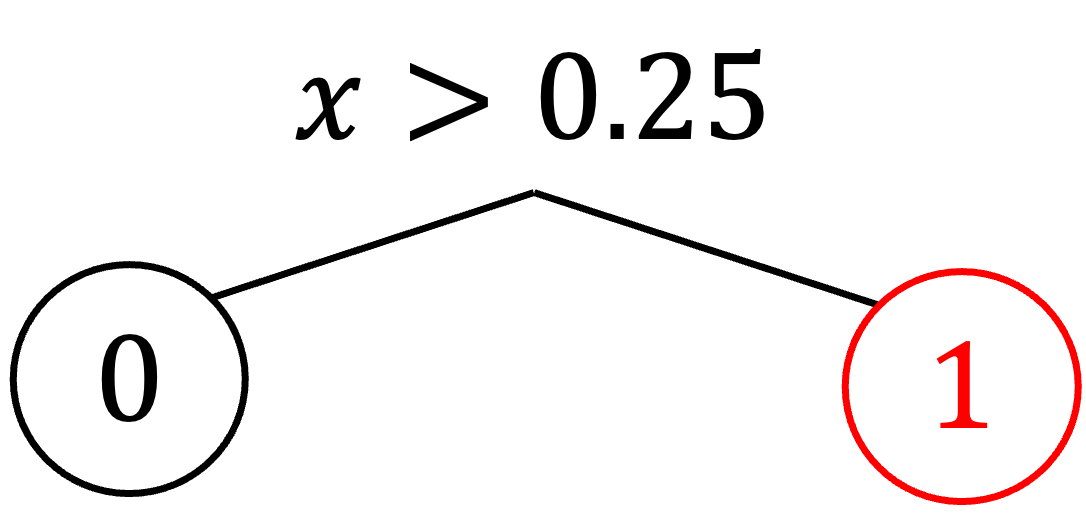}
\end{minipage}
\caption{{\em Left:} An instance of $k(x;W)$ with $k_{\min} = 0$ for $W = [0.25, 0.5]$ (black), with a proposed shift in the second wrapping number (red).  The tree diagram below it represents the equivalent partition of the input space, with the change in splitting criteria in red. {\em Middle:} Proposal of a new wrapping location, equivalent to adding a split in the tree. {\em Right:} Removal of one of the wrapping locations, equivalent to pruning the tree.}
\label{fig:tree}
\end{figure}

To properly explore the space of wrapping numbers $K$, we propose three types of changes to $W$.  These are shown in Figure \ref{fig:tree}.  First we consider taking an existing wrapping location $w_i$ and ``shifting'' it to the left or right, which is equivalent to changing one of the splitting criteria (left panel).  This will either increase or decrease the wrapping numbers of observations contained in the region between the old and new wrapping locations by one.  We also propose ``growing'' a new wrapping location, equivalent to adding a new split $w_{m+1}$ and increasing the depth of the partitioning tree (middle panel).  Observations with values of $x$ larger than this new split point will have their wrapping number increase by one.  Conversely, we also propose ``shrinking'' one of the existing wrapping locations, which can be achieved by pruning one of the partitions (right panel).  These three types of movement maintain monotonicity in $K$.

\begin{algorithm}[ht!]
\DontPrintSemicolon
{\bf Input:} Current $W^{(t)} = \{w_1^{(t)},\dots,w_m^{(t)}\}$, $K^{(t)}$, response $Y$, likelihood $\mathcal{L}$ \\
{\bf Output:} Wrapping locations $W^{(t+1)}$, Posterior sample $K^{(t+1)} \sim p(K\mid Z, Y)$ \\
~
Set $W^* = W^{(t)}$, $w_0 = x_{\min}, w_{m+1} = x_{\max}$\\
\For{$i \in 1,\dots,m$ \tcp*{Shift}}{
Draw $w^*\sim \text{Unif}(w_{i-1},w_{i+1})$\\
Set $W_i^* = w^*$, propose $K^* = k(X;W^*)$\\
Accept $W^{(t+1)} = W^*$, $K^{(t+1)} = K^*$ with probability $r = \min\left(1, \dfrac{\mathcal{L}\left(Z,K^*\mid Y\right)}{\mathcal{L}\left(Z,K^{(t)}\mid Y\right)}\right)$}

Draw $w^*\sim\text{Unif}(x_{\text{min}}, x_{\text{max}})$ \tcp*{Grow}
Set $W^* = W^* \cup w^*$ such that $w^*$ is sorted in ascending order, propose $K^* = k(X; W^*)$\\
Accept $W^{(t+1)} = W^*$, $K^{(t+1)} = K^*$ with probability $r = \min\left(1, \dfrac{\mathcal{L}\left(Z,K^*\mid Y\right)}{\mathcal{L}\left(Z,K^{(t)}\mid Y\right)}\cdot\dfrac{x_{\max}-x_{\min}}{m+1}\right)$

Set $m = |W^{(t+1)}|$ \tcp*{Shrink}
Draw $i\sim \text{Unif}\{1,\dots,m\}$\\
Set $W^* = W^{(t+1)}\setminus w_i$, propose $K^* = k(X;W^*)$\\
Accept $W^{(t+1)} = W^*$, $K^{(t+1)} = K^*$ with probability $r = \min\left(1, \dfrac{\mathcal{L}\left(Z,K^*\mid Y\right)}{\mathcal{L}\left(Z,K^{(t)}\mid Y\right)}\cdot\dfrac{m}{x_{\max}-x_{\min}}\right)$\\
Return $W^{(t+1)}$, $K^{(t+1)}$
\caption{Metropolis-Hastings procedure for new wrapping numbers}
\label{alg:k_samp}
\end{algorithm}

Our procedure for posterior sampling from $K\mid Z, Y$ through new partitioning functions $k(X;W)$ is outlined in Algorithm \ref{alg:k_samp}.  Each type of movement described in Figure \ref{fig:tree} is proposed in a single MCMC iteration.  Shifting entails iterating through each current wrapping location $w_i$ and moving it uniformly at random between $w_{i-1}$ and $w_{i+1}$.  Note the minimum and maximum values of $w$ are fixed at $x_{\min}$ and $x_{\max}$; that is, we don't consider wrapping beyond the range of observed inputs.  The proposal of a new partitioning function $W^*$ elicits a proposed set of wrapping numbers $K^*$, which is evaluated with a likelihood ratio.  Next, an additional wrapping location $w^*$ is proposed uniformly at random in the standardized input space.  After updating $W^*$ with $w^*$, $K^*$ is reconstructed accordingly.  Proposed shrinking selects one of the existing wrapping locations at random to remove from $W^*$.  Note the terms $(x_{\max}-x_{\min})/{(m+1)}$ and $m/(x_{\max}-x_{\min})$ in the acceptance ratios for growing and shrinking, respectively, stem from the proposal functions which are inherently inverses of one another; a growth can be reversed by selecting that wrapping number to remove with probability $1 / (m+1)$, and a shrink can be reversed by adding a wrapping point at that particular location with probability $1 / (x_{\max}-x_{\min})$.

Although we save a complete discussion of prediction for Section \ref{ss:wrap_fit}, we mention now that inducing wrapping numbers through the evaluation of a partitioning function makes inference of wrapping numbers for novel inputs straightforward.  With posterior samples $W^{(1)},\dots,W^{(T)}$ obtained through Algorithm \ref{alg:k_samp}, we can estimate properties of the wrapping number $k'$ for a new input $x'$ via MC approximation:
\begin{align}
\hat{\mu}(k') &= \dfrac{1}{T}\sum_{t=1}^T k(x'; W^{(t)}) & \hat{\sigma}^2(k') &= \dfrac{1}{T}\sum_{t=1}^T\left(k(x';W^{(t)}) - \hat{\mu}(k')\right)^2.
\label{eq:k_pred}
\end{align}

\begin{figure}[ht!]
\centering
\includegraphics[scale=0.55, trim=0 15 0 50, clip=TRUE]{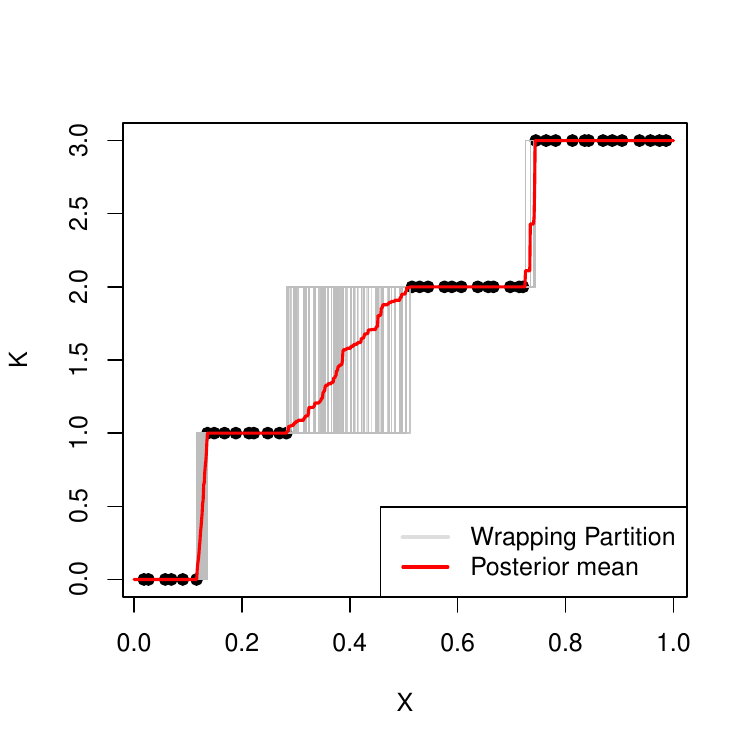}
\includegraphics[scale=0.55, trim=0 15 0 50, clip=TRUE]{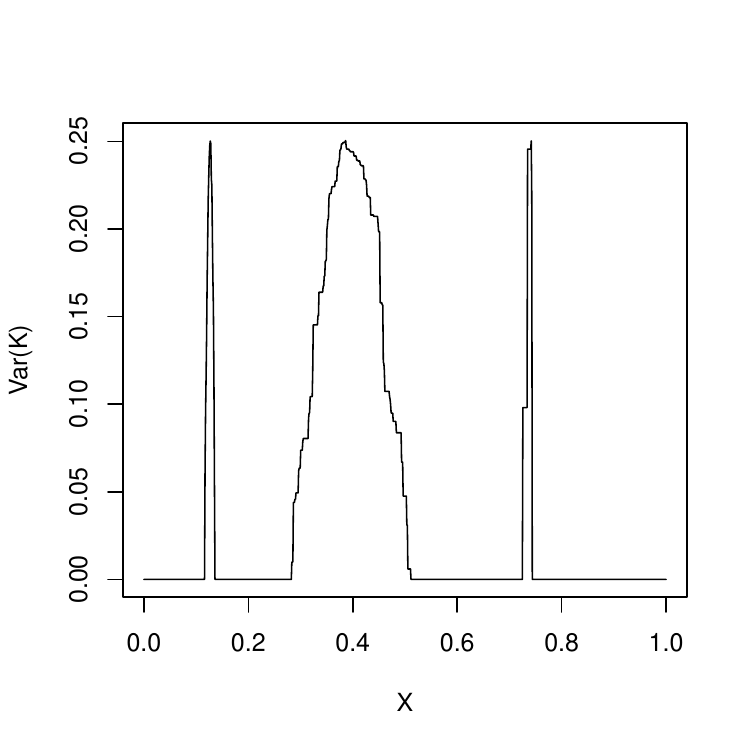}
\caption{{\em Left:} Posterior samples of the wrapping number partitions, along with the posterior mean estimate. {\em Right:} Posterior predictive variance for $K$ evaluated on fine test grid.}
\label{fig:k_est}
\end{figure}
Figure \ref{fig:k_est} illustrates our algorithm's performance on the same example presented earlier in Figure \ref{fig:missing}.  We ran our MH procedure for $10{,}000$ iterations and include posterior samples of the wrapping locations $W$ after a burn-in of $5{,}000$ and keeping every $10^{\text{th}}$ iteration.  Note we estimate a jump in wrapping numbers with roughly equal mass across input values ranging from $0.2$ to $0.5$, matching our intuition from Figure \ref{fig:missing}.  We also plot the predictive mean (\textit{left}) and variance (\textit{right}) for $K$ on a fine grid of inputs of size $1{,}000$ in the left and right panels, calculated using Eq.~(\ref{eq:k_pred}).  Peaks are observed where the function wraps at around $x=0.15$ and $x=0.75$, and in the gap of missing data, where uncertainty is maximized at the point farthest away from observed inputs.

\subsection{Latent GP Inference} \label{ss:ess}
The specification of a Student's $t$ likelihood in Eq.~(\ref{eq:like}) eliminates prior conjugacy for the unwrapped GP $Z$ and requires other means of posterior estimation.  Rather than rely on rejection sampling or MH approaches, we can take advantage of elliptical slice sampling \citep[ESS;][]{murray2010elliptical} to perform rejection-free sampling from the posterior for $Z$.  ESS has been used to infer latent outputs in generalized GP settings with non-Gaussian likelihoods \citep{barnett2025monotonic}, as well as sampling from inner layers of ``deep'' GPs \citep{sauer2023active}.  The sampling framework crucially hinges on the Gaussian prior form of $Z$ in Eq.~(\ref{eq:wrap_model}).  Given a valid sample posterior $Z^{(t)}$, and a single (de-meaned) prior GP instance $Z'\sim \mathcal{N}\left(0, \tau^2\Sigma_\theta(X)\right)$, we can identify an $n$-dimensional elliptical ``slice'' which both of these quantities lie on.  We can then identify a valid posterior sample on that slice through iterative tuning.  

\begin{algorithm}[ht!]
\DontPrintSemicolon
{\bf Input:} Previous $Z^{(t)}, K^{(t)}$, training data $X, Y$, covariance $\tau^2\Sigma_\theta(X)$, log-likelihood $\ell$ \\
{\bf Output:} Posterior sample $Z^{(t+1)} \sim p(Z\mid Y,K)$ \\
~

Draw $Z' \sim \mathcal{N}_n\left(0,\tau^2\Sigma_\theta(X)\right)$ \;
Draw $u \sim \text{Unif}[0, 1]$ and set acceptance threshold $\ell_{\text{thresh}} = \ell\left(Z^{(t)},K^{(t)}\mid Y\right) + \log(u)$ \;
Draw angle $\gamma \sim \text{Unif}[0,2\pi]$ and construct bracket $\gamma_{\text{min}} = \gamma - 2\pi$, $\gamma_{\text{max}} = \gamma$ \;
~

\While{\text{True}}{
Calculate proposal $Z^* = Z^{(t)}\cos(\gamma) + (Z' + \alpha + \beta X)\sin(\gamma)$ and evaluate $\ell_{\text{prop}} = \ell(Z^*,K^{(t)}\mid Y)$ \;
\If{$\ell_{\text{prop}} > \ell_{\text{thresh}}$}{Return $Z^{(t+1)} = Z^*$}
\Else{
\If{$\gamma < 0$}{$\gamma_{\text{min}}=\gamma$}
\Else{$\gamma_{\text{max}}=\gamma$}
Draw new $\gamma \sim \text{Unif}[\gamma_{\text{min}}, \gamma_{\text{max}}]$
}
}
\caption{Elliptical slice sampling for WGP estimation}
\label{alg:ess}
\end{algorithm}
Our ESS procedure is outlined in Algorithm \ref{alg:ess}.  Proposals are evaluated based on the log-likelihood for $Y$ given in Eq.~(\ref{eq:like}).  What's crucial about this scheme is that each ESS call requires only one GP prior draw; further proposal refining only requires narrowing down an appropriate angle on the ellipse $\gamma$.  This means ESS is computationally cheaper than a rejection sampling approach, since the most expensive operation required for proposals is sampling from an MVN.  This is due to needing to invert an $n\times n$ covariance matrix $\Sigma_\theta(X)$, which if done using a Cholesky decomposition runs in $\mathcal{O}(n^3)$ time. 

\begin{figure}[ht!]
\centering
\includegraphics[width=0.32\textwidth, trim=0 10 0 40, clip=TRUE]{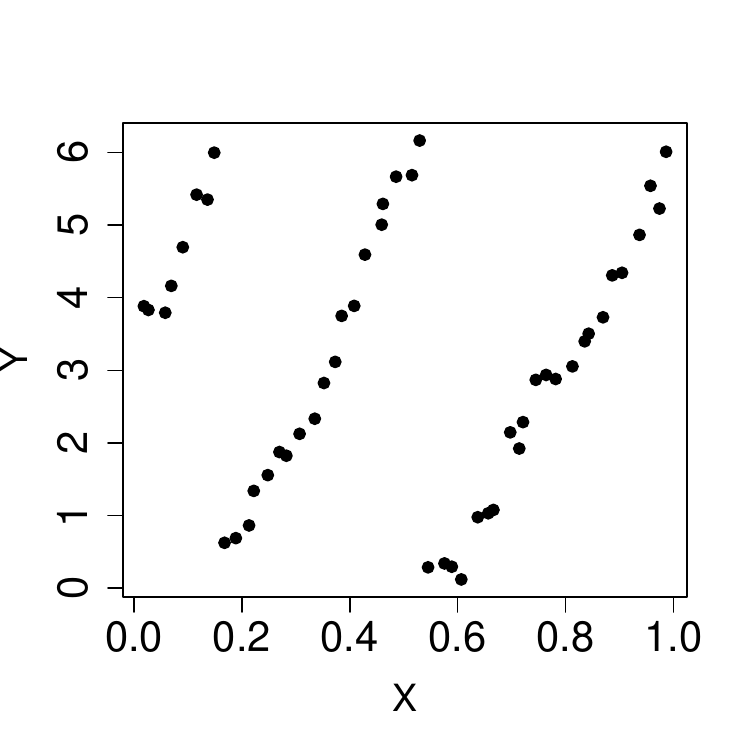}
\includegraphics[width=0.32\textwidth, trim=0 10 0 40, clip=TRUE]{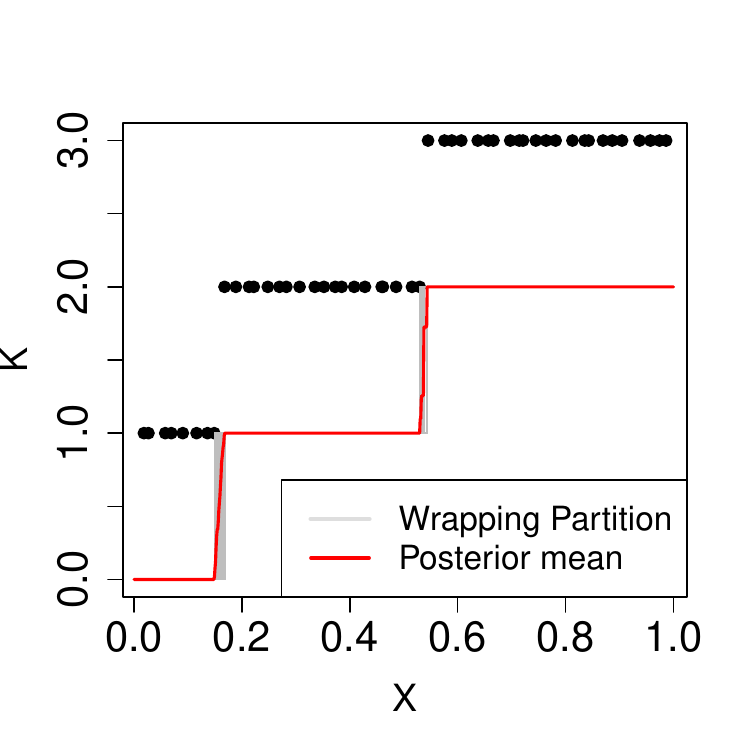}
\includegraphics[width=0.32\textwidth, trim=0 10 0 40, clip=TRUE]{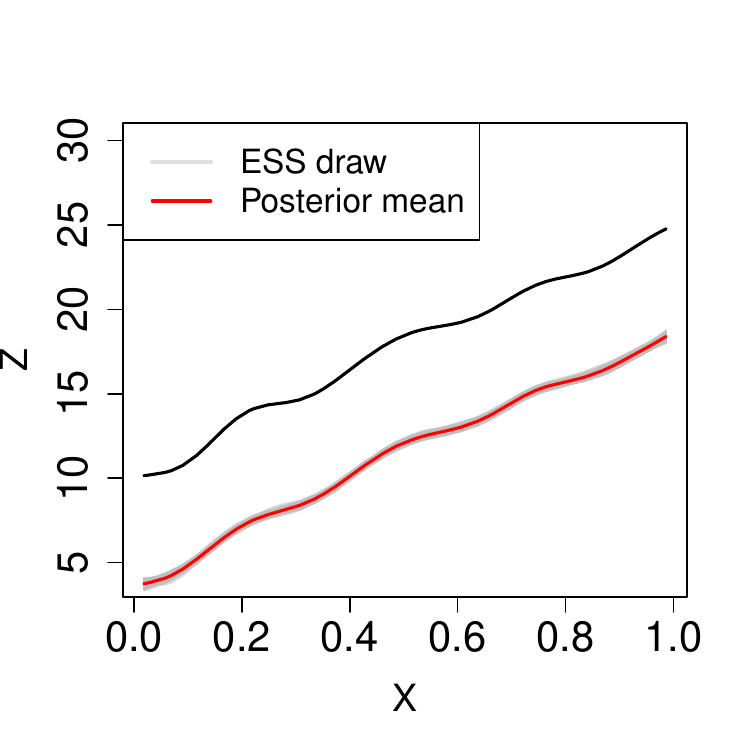}
\caption{{\em Left:} Simulated angular data of size $n=50$ generated from a linear mean wrapped GP instance with independent Student's $t$ distributed noise added.  {\rm Middle:} The true known wrapping numbers $k_1,\dots,k_n$ for each observation, along with sampled wrapping partitions from our MCMC algorithm and the posterior mean estimate included in red.  {\rm Right:} Posterior draws from the latent unwrapped GP sampled using ESS.  The truth is shown in black.}
\label{fig:ess_ex}
\end{figure}

We showcase ESS performance for wrapped GP fitting on a simulated example, seen in Figure \ref{fig:ess_ex}.  A Latin hypercube sampling (LHS) design $X$ of size $n=50$ was generated for training. The left panel shows angular responses $Y$ produced from a linear-mean wrapped GP model with settings $\alpha = 10$, $\beta = 20$, $\theta = 0.01$, $\tau^2 = 1$, $\sigma^2 = 0.05$, and $\nu = 5$.  The true, known wrapping numbers obtained from wrapping around the unit circle are shown in the middle panel, along with samples of the wrapping number partition generated from Algorithm \ref{alg:k_samp}.  In the right panel we plot a chain of $500$ posterior samples ($10{,}000$ in total minus a burn-in of $5{,}000$) from $Z\mid K,Y$, generated via Algorithm \ref{alg:ess}, with the true function shown in black.  Note our estimations of both $K$ and $Z$ underestimate the true wrapping numbers and unwrapped GP by factors of $1$ and $2\pi$, respectively.  This stems from an inherent identifiability issue in wrapped settings; equivalent data could be generated with an instance of $Z$ shifted down by $2\pi$ and a value of $K$ that starts at $0$, for instance.  As we show in Section \ref{ss:wrap_fit}, this discrepancy between our estimation of $K$ and $Z$ and the truth does not impact our ability to perform accurate out-of-sample prediction.

\subsection{Wrapped Gaussian Process Estimation} \label{ss:wrap_fit}
We now combine the algorithms proposed in Sections \ref{ss:wrap_num} and \ref{ss:ess} into a single MCMC procedure for estimation of our WGP model, in which parameters are sampled iteratively from their corresponding conditional posterior distributions in a Gibbs scheme.  
\begin{algorithm}[ht!]
\DontPrintSemicolon
{\bf Input:} Starting values for $W^{(1)}, K^{(1)}, Z^{(1)}, \alpha^{(1)}, \beta^{(1)}, \tau^{2(1)}, \theta^{(1)}, \sigma^{2(1)}, \nu^{(1)}$\\
{\bf Output:} $T$ posterior samples from $W, K, Z, \alpha, \beta, \tau^2, \theta, \sigma^2, \nu$\\
~
\For{$t \in 1,\dots,T$}{
Draw $(W,K)^{(t+1)} \sim K\mid Y,W^{(t)},K^{(t)},Z^{(t)},\alpha^{(t)},\beta^{(t)},\tau^{2(t)},\theta^{(t)},\sigma^{2(t)},\nu^{(t)}$ \tcp*{Alg.~(\ref{alg:k_samp})}
Draw $Z^{(t+1)}\sim Z\mid Y,Z^{(t)},K^{(t+1)},\alpha^{(t)},\beta^{(t)},\tau^{2(t)},\theta^{(t)},\sigma^{2(t)},\nu^{(t)}$ \tcp*{Alg.~(\ref{alg:ess})}
Draw $(\alpha, \beta)^{(t+1)}\sim (\alpha,\beta)\mid Y,Z^{(t+1)},\tau^{2(t)},\theta^{(t)}$ \tcp*{Gibbs}
Draw $\tau^{2(t+1)}\sim \tau^2\mid Y,Z^{(t+1)},\alpha^{(t+1)},\beta^{(t+1)},\theta^{(t)}$ \tcp*{Gibbs}
Draw $\theta^{(t+1)}\sim \theta\mid Y,\theta^{(t)},Z^{(t+1)},\alpha^{(t+1)},\beta^{(t+1)},\tau^{2(t+1)}$ \tcp*{MH}
Draw $\sigma^{2(t+1)}\sim \sigma^{2}\mid Y,Z^{(t+1)},K^{(t+1)},\nu^{(t)}$ \tcp*{MH}
Draw $\nu^{(t+1)}\sim \nu\mid Y,Z^{(t+1)},K^{(t+1)},\sigma^{2(t+1)}$ \tcp*{MH}}
\caption{MCMC procedure for WGP estimation}
\label{alg:mcmc}
\end{algorithm}
Our approach is outlined in Algorithm \ref{alg:mcmc}.  In addition to $K$ and $Z$ we sample from mean parameters $\alpha$ and $\beta$, covariance hyperparameters $\tau^2$ and $\theta$, and likelihood parameters $\sigma^2$ and $\nu$.  For the mean and intercept we specify a joint prior $\alpha, \beta \sim \mathcal{N}_2\left(\begin{bmatrix}0, \mu_0\end{bmatrix}^\top, \sigma^2_0I_{2\times 2}\right)$.  This yields a conjugate posterior form $\alpha, \beta \mid Z,\tau^2,\theta \sim \mathcal{N}_2\left(A^{-1}b, A^{-1}\right)$, where
\begin{align*}
        A &= \dfrac{1}{\tau^2}\left(X^\top \Sigma_n^{-1}X + \dfrac{1}{\sigma_0^2}I_2\right) & 
        b &=  \dfrac{1}{\tau^2}\left(X^\top \Sigma_n^{-1}Z + \sigma_0^2\begin{bmatrix}0 \\ \mu_0\end{bmatrix}\right).
\end{align*}

As mentioned earlier in Section \ref{ss:wrap_num}, the slope strongly influences the wrapping behavior of $Z$, in that a larger $\beta$ creates more wrapping.  For this reason, we have found the need to provide an informative value for $\mu_0$.  We specify the prior mean with a simple local linear approximation as a pre-processing step.  Given $X$ is sorted in ascending order, we partition observations into contiguous groups $c_1,\dots,c_{\lceil n/\ell \rceil}$, each of size $\ell < n$.  For each subset $c_j$ we estimate the local slope $\hat{\beta}_j = \sum_{i\in c_j} (x_i - \bar{x}_{c_j})(y_i - \bar{y}_{c_j}) / \sum_{i\in c_j} (x_i - \bar{x}_{c_j})^2$; we then take the empirical average of all of the slope estimates lower-bounded by zero: $\mu_0 = \sum_{j=1}^{\lceil n/\ell \rceil} \max\{0, \hat{\beta}_j\} / \lceil n/\ell\rceil$.  We specify a prior slope variance $\sigma_0^2 = 10$.

For hyperparameters $\tau^2$ and $\theta$ we specify $\text{IGa}(1,1)$ and $\text{Ga}(5/2,3/2)$ priors, respectively.  We sample the scale parameter from its posterior form: $\tau^2\mid Z,\alpha,\beta,\theta \sim \text{IGa}\left(a_n, b_n\right)$, where $a_n = (1 + n)/2$ and $b_n = \left(1 + (Z-\alpha-\beta X)^\top(Z - \alpha - \beta X)\right)/2$.  For the lengthscale we must employ an MH procedure due to non-conjugacy.  For the likelihood we specify priors for the common variance $\sigma^2 \sim \text{Ga}\left(1/2,1/2\right)$ and degrees of freedom $\nu \sim \text{Exp}(1/30)$, and use separate MH steps to sample from their posteriors.  Note the continuous distribution for $\nu$ allows for non-whole values.  In order for the variance to be finite, the degrees of freedom associated with the Student's $t$ distribution must be greater than $2$; we therefore enforce a prior restriction that $\nu \geq 3$.

A common objective of GP regression is to perform prediction at unobserved input locations.  Conditional on training data $X$ and $Y$, we wish to derive the marginal predictive distribution for $\mathcal{Y}$ corresponding to $n'$ new inputs $\mathcal{X}\in \mathbb{R}^{n'}$.  In addition to $Z$ and $K$, we must integrate out and sum over the latent unwrapped output $\mathcal{Z}$ and wrapping numbers $\mathcal{K}$ corresponding to $\mathcal{X}$, respectively.  Omitting mean, variance and covariance parameters for brevity, this quantity can be expressed as
\begin{align*}
    p(\mathcal{Y}\mid X,Y) &= \int_{\mathcal{Z}\in\mathbb{R}^{n'}}\sum_{\mathcal{K}\in\mathbb{Z}^{n'}} p(\mathcal{Y}\mid\mathcal{Z},\mathcal{K})p(\mathcal{Z},\mathcal{K}\mid X,Y)d\mathcal{Z}\\
    &= \int_{\mathcal{Z}\in\mathbb{R}^{n'}}\sum_{\mathcal{K}\in\mathbb{Z}^{n'}}\int_{Z\in\mathbb{R}^n} \sum_{K\in\mathbb{Z}^n}p(\mathcal{Y}\mid \mathcal{Z},\mathcal{K})p(\mathcal{Z},\mathcal{K}\mid X,Z,K)p(Z,K\mid X,Y)dZd\mathcal{Z}.
\end{align*}
Because of the intractability of its PDF, we approximate properties of $\mathcal{Y}$ via MC.  Samples from $p(\mathcal{Z},\mathcal{K}\mid X,Z,K) = p(\mathcal{Z}\mid X,Z)p(\mathcal{K}\mid K)$ are obtained by drawing new $\mathcal{Z}$ and $\mathcal{K}$ separately.  As shown earlier in Section \ref{ss:wrap_num}, draws from $\mathcal{K}$ can be obtained by evaluating the partitioning function $k(\mathcal{X};W^{(t)})$ that accompanies each sample $K^{(t)}$.  Instances of $\mathcal{Z}$ can be drawn through the well-known ``kriging'' equations \citep{cressie1988spatial}.  Let $\Sigma_\theta(\mathcal{X})$ and $\Sigma_\theta(\mathcal{X},X)$ represent $n'\times n'$ and $n'\times n$ matrices, respectively, that are constructed analogously to $\tau^2\Sigma_\theta(X)$ using $\tau^2$ and $\theta$.  Then $\mathcal{Z}\mid X, Z,\alpha,\beta,\tau^2,\theta \sim \mathcal{N}_{n'}(\mu_\mathcal{X}, \Sigma_\mathcal{X})$, where
\begin{equation}
\begin{aligned}
        \mu_\mathcal{X} &=  \alpha 1_{n'} + \beta\mathcal{X} + \Sigma_\theta(\mathcal{X},X)\Sigma^{-1}_\theta(X)(Z - \alpha 1_{n'} - \beta X)\hspace{5mm} \\
        \Sigma_\mathcal{X} &= \tau^2\Sigma_\theta(\mathcal{X}) - \tau^2\Sigma_\theta(\mathcal{X},X)\Sigma_\theta^{-1}(X)\Sigma_\theta(\mathcal{X},X)^\top.
\end{aligned}
\label{eq:krig}
\end{equation}
We can generate $\mathcal{Z}^{(1)},\dots,\mathcal{Z}^{(T)}$ by taking posterior samples obtained through Algorithm \ref{alg:mcmc}, feeding them through Eq.~(\ref{eq:krig}), and drawing from the MVN.  With both samples $(\mathcal{Z},\mathcal{K})^{(1)},\dots,(\mathcal{Z},\mathcal{K})^{T}$, as well as  $(\sigma^2,\nu)^{(1)},\dots,(\sigma^2,\nu)^{(T)}$ obtained from Algorithm \ref{alg:mcmc}, we can approximate the marginal predictive mean and variance for each $y_i\in\mathcal{Y}$ as
\begin{equation}
\begin{aligned}
        \hat{\mu}_{y_i} &= \dfrac{1}{T}\sum_{t=1}^T \left(\mathcal{Z}_i^{(t)} - 2\pi\mathcal{K}_i^{(t)}\right) \\
        \hat{\sigma}^2_{y_i} &= \dfrac{1}{T}\sum_{t=1}^T \sigma^{2(t)}\dfrac{\nu^{(t)}}{\nu^{(t)}-2} + \dfrac{1}{T-1}\sum_{t=1}^T\left(\mathcal{Z}_i^{(t)} - 2\pi\mathcal{K}_i^{(t)} - \hat{\mu}_{y_i}\right)^2.
\end{aligned}
\label{eq:pred_eqs}
\end{equation}
The first term in the predictive variance in Eq.~(\ref{eq:pred_eqs}) estimates the marginal variance for a Student's $t$ distributed random variable; the second quantity estimates the variance of our model's estimation of the predictive mean.  Joint sampling from $\mathcal{Z}$ is preferred because it accounts for pairwise covariances between outputs, but sampling from an MVN distribution can be costly for large $n$.  To save time without significantly sacrificing predictive performance, we opt for pointwise prediction instead, where the posterior predictive for each $z\in\mathcal{Z}$ simplifies to a univariate Gaussian, and the parameters in Eq.~(\ref{eq:krig}) reduce to scalars.

\begin{figure}[ht!]
\centering
\includegraphics[scale=0.55, trim=0 10 0 50, clip=TRUE]{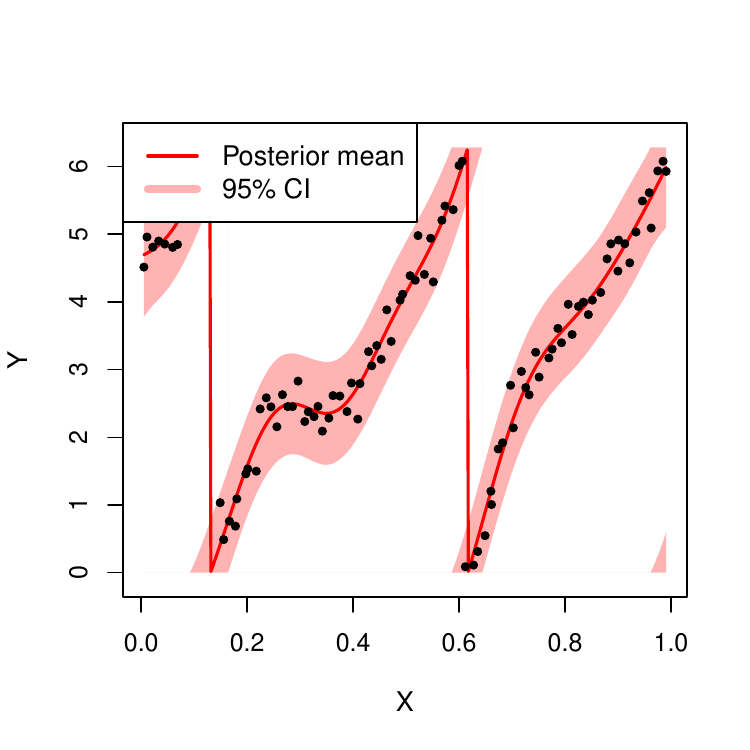}
\includegraphics[scale=0.55, trim=0 10 0 50, clip=TRUE]{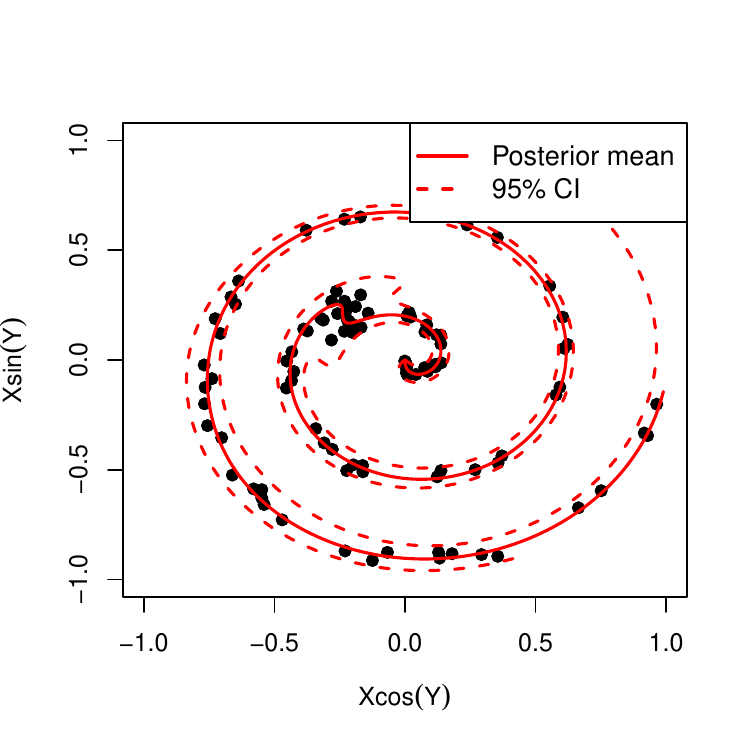}
\caption{{\em Left:} Posterior mean predictions and $95\%$ credible interval from our WGP approach.  {\em Right:} The same data and predictions transformed into polar coordinates.}
\label{fig:ex_fit}
\end{figure}
We showcase predictive performance of our WGP model on the same simulated example first introduced in Section \ref{ss:gp_reg}, illustrated in Figure \ref{fig:ex_fit}.  The left panel shows posterior mean predictions and a $95\%$ credible interval from a WGP model trained on an LHS design of size $n=100$, and predicted on a fine grid of size $n' = 500$.  The right panel plots data in the polar plane via trigonometric identities: $\left(X\cos(Y),X\sin(Y)\right)$.  Compared to the ordinary GP, our model's appropriate capturing of the wrapping behavior leads to more appropriate UQ within the range of the training data.  Also note that rather than expand beyond the domain of the response, the uncertainty interval wraps around the unit circle along with the mean.

\section{Results on Synthetic Examples} \label{sec:sim}
In this Section we apply our WGP model to simulated data from functions that exhibit monotonic wrapping behavior.  We compare our method with two other WGP implementations: the original WGP formulation proposed by \cite{jona2012spatial}, and the manifold learning approach proposed by \cite{mallasto2018wrapped}.  We implement the former in \texttt{R} ourselves since we could not find code available online.  For the later we use the Python implementation available in the ``geomstats'' library \citep{geomstats}.  Code to use our method and run these experiments in \texttt{R} are available online\footnote{\url{https://github.com/lanl/wrapgp}}.  

For each simulated example we generate $10$ separate training data sets using LHS designs of increasing sizes.  We evaluate each model's predictive performance on the same hold-out testing sets generated from the same $10$ LHS designs of size $n' = 500$, from which we estimate pointwise mean $\hat{\mu}_i$ and standard deviation $\hat{\sigma}_i$ for $i=1,\dots,n'$.  For metrics we use a variation of root-mean square error (RMSE) appropriate for angular predictions, which properly calculates ``circular'' residuals:
\begin{align*}
    \text{RMSE}(\mathcal{Y},\hat{\mu}) &\equiv \dfrac{1}{n'}\sum_{i=1}^{n'}r(y_i, \hat{\mu}_i)^2,
    && \text{where}  &
    r(a, b) &= \underset{\ell\in\{-1,0,1\}}{\min}\left\{\left|a + 2\pi\ell - b\right|\right\}.
\end{align*}
Circular residuals can be no greater than $\pi$, which occurs when an observation and its prediction lie at opposite ends of the circle.  

To assess UQ we calculate the continuous ranked probability score (CRPS).  CRPS evaluates probabilistic forecasts based on the cumulative density function (CDF) of the response.  Since our comparators assume a different likelihood, we approximate CRPS based on a quantile decomposition using posterior predictive samples from $\mathcal{Y}$ \citep{laio2007verification}:
\begin{align*}
\text{CRPS}(\mathcal{Y},\hat{\mathcal{Y}}) &\equiv \dfrac{2}{n'T}\sum_{i=1}^{n'}\sum_{t=1}^T\left(\mathbbm{1}\left(y_i < \hat{y}_i^{(t)}\right)-s_i\right)\left(\hat{y}_i^{(t)}-y_i\right), \\ \text{ where} \quad\quad
s &= \left[\dfrac{1}{2T},\dots,1-\dfrac{1}{2T}\right]
\end{align*}
is an evenly spaced sequence of size $T$.  To account for predictions for observations that wrap around, we set each sample's pointwise prediction $\hat{y}^{(t)} = y^{(t)} + 2\pi l$, where 
\begin{align*}
l = \underset{\ell\in\{-1,0,1\}}{\argmin}\left\{\left|y^{(t)} + 2\pi\ell - y\right|\right\}.
\end{align*}
This allows predictions to shift below $0$ or above $2\pi$ to reflect closer residuals than what appear in circular space.

\subsubsection*{Logarithmic Function} We use the simple logarithmic function from \cite{holsclaw2013gaussian} and cataloged in Simon Fraser University's Virtual Library of Simulation Experiments.\footnote{\url{https://www.sfu.ca/~ssurjano/holsetal13log.html}}  We scale the output by a factor of $15$ to increase the amount of wrapping observed: $z = 15\log(1 + x)$.  We also take inputs from our LHS designs and multiply them by a factor of $2.5$ to ensure $x\in[0,2.5]$.  Like \cite{holsclaw2013gaussian} we add independent, Gaussian distributed error $\epsilon \sim \mathcal{N}(0,0.5^2)$ to the output, after which we take modulo-2$\pi$ to ensure responses are valid angles: $y = (z + \epsilon)\mod 2\pi$.   
\begin{figure}[ht!]
\centering
\includegraphics[width=0.4\textwidth,trim=20 20 0 20]{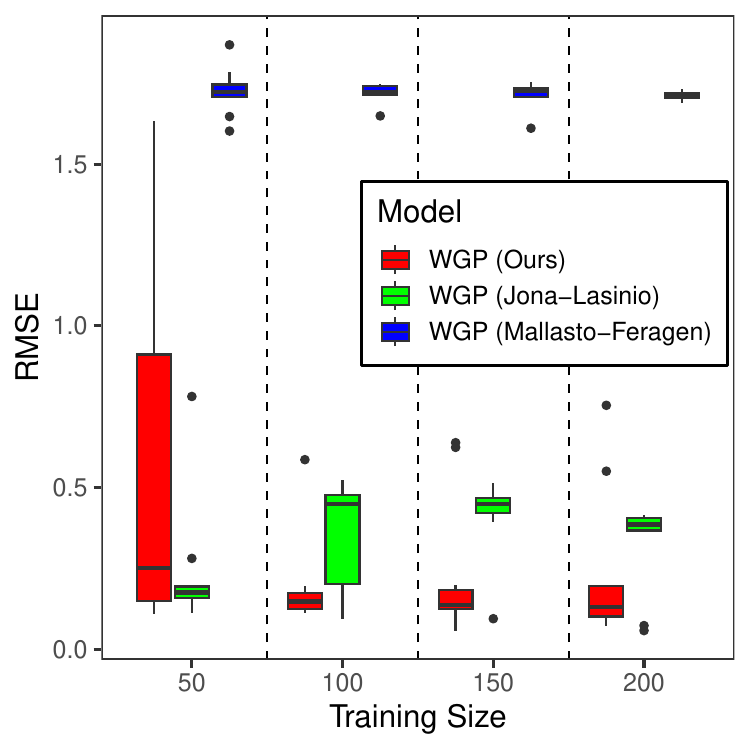}
\hspace{1cm}
\includegraphics[width=0.4\textwidth,trim=20 20 0 20]{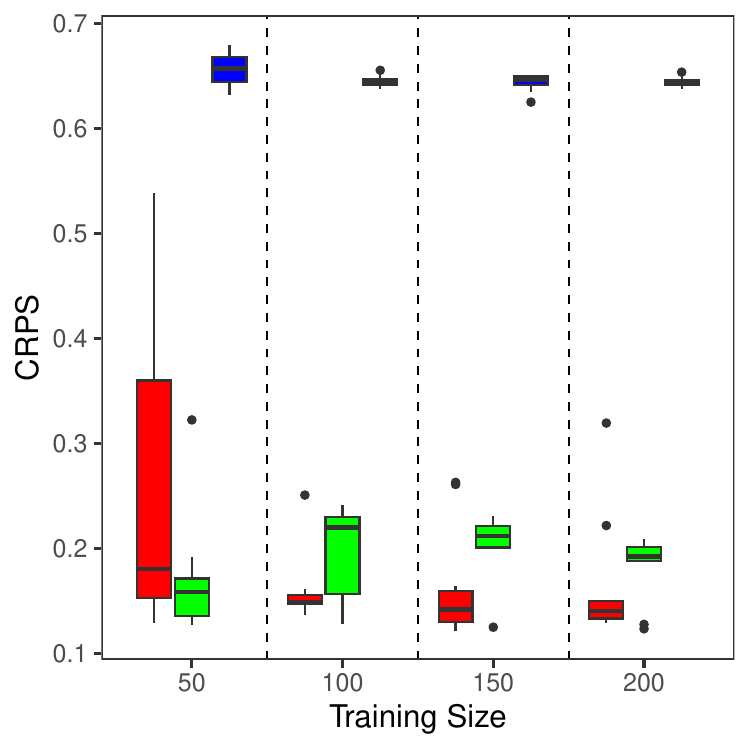}
\caption{Results for the logarithmic function example.  Boxplots summarize out-of-sample performance with circular RMSE (left) and CRPS (right) for all three WGP methods across $10$ random training sets of increasing size.  Smaller values indicate higher accuracy.}
\label{fig:log_results}
\end{figure}
Results from the simulation experiment are given in Figure \ref{fig:log_results}.  Poorer performance with training sizes of $n=50$ stems from missing one of the wrapping locations, often due to underestimating the slope.  For all other training sizes, our WGP implementation achieves the smallest median RMSE and exhibits the smallest variability in performance across training sizes.  Our method also achieves the smallest median CRPS for $n \geq 100$.  The WGP from Mallasto and Feragen performed consistently worst due to failing to capture the wrapping behavior of the function.  

\subsubsection*{Logarithmic Function with Input Gaps} We now slightly modify the logarithmic example by only sampling the function on a subset of the unit interval.
\begin{figure}[ht!]
\centering
\includegraphics[scale=0.55,trim=20 30 0 60]{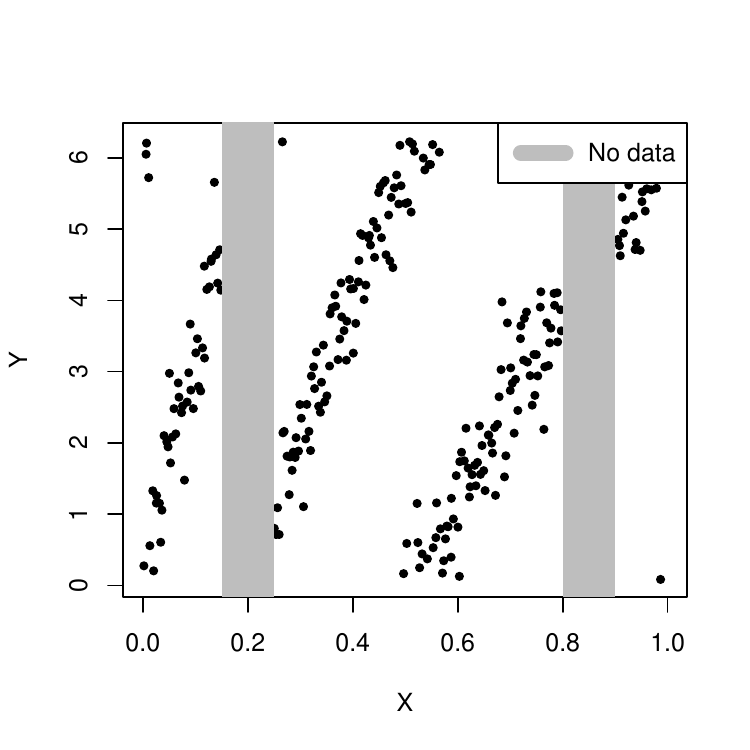}
\hspace{1cm}
\includegraphics[scale=0.55,trim=20 30 0 60]{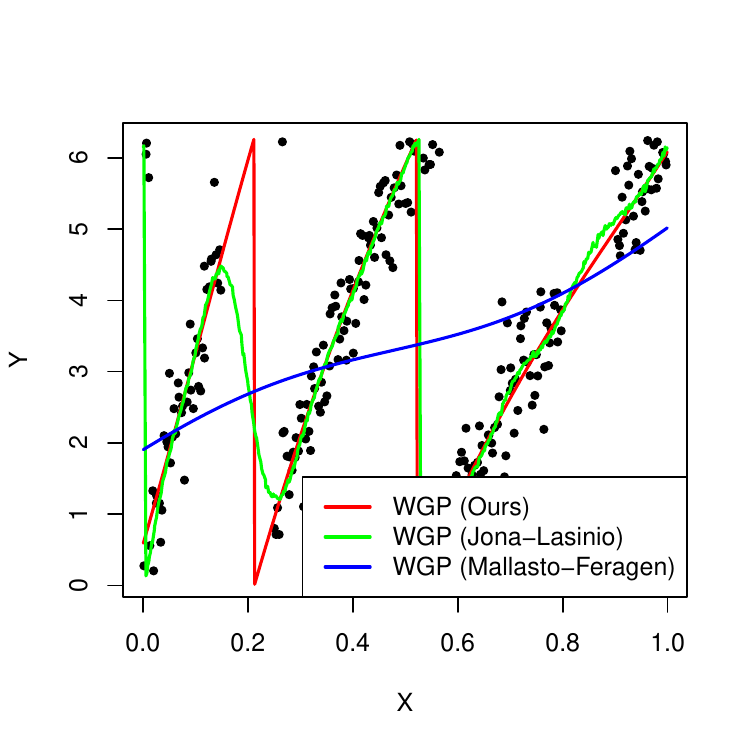}
\caption{{\em Left:} Gaps in logarithmic function. {\em Right:} Comparison of mean predictions from the three WGP implementations.}
\label{fig:log_gap_ex}
\end{figure}
The left panel of Figure \ref{fig:log_gap_ex} illustrates two regions of the input space where no data is sampled in the gray boxes.  We purposely censor regions where wrapping is known to occur (left box) and a region where no wrapping occurs (right box).  Separate LHS designs were generated for the three strata $[0,0.15], [0.25,0.8], [0.9,0.1]$ with sizes proportional to their interval widths, such that their total summed up to the training size (in this example $n = 200$).  In the right panel we visually compare predictive performance on a separate (uncensored) LHS design across the three WGP implementations.  While our method identifies all of the wrapping locations, Jona-Lasinio misses the wrapping around $0.2$, and Mallasto and Feragen misses both.

\begin{figure}[ht!]
\centering
\includegraphics[width=0.4\textwidth,trim=20 0 0 0]{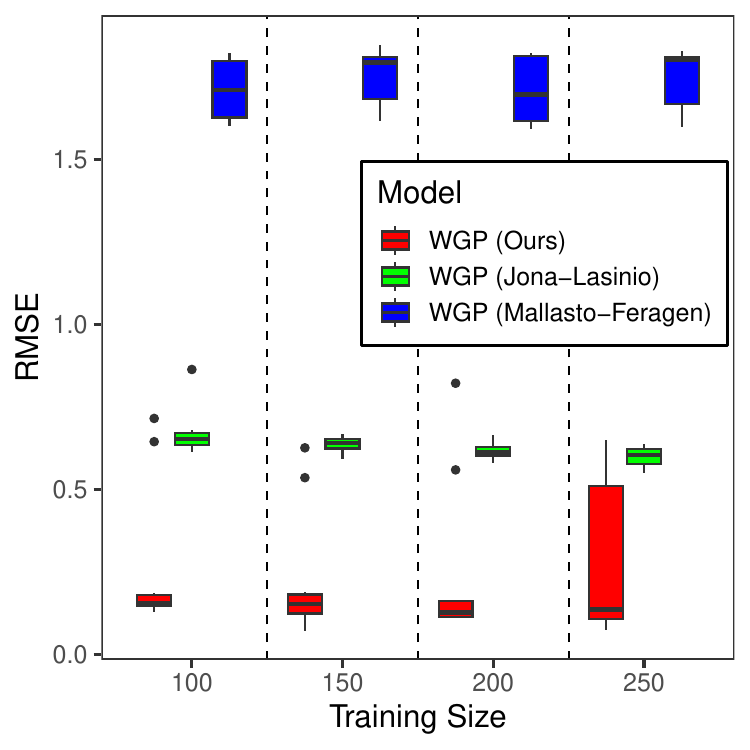}
\hspace{1cm}
\includegraphics[width=0.4\textwidth,trim=20 0 0 0]{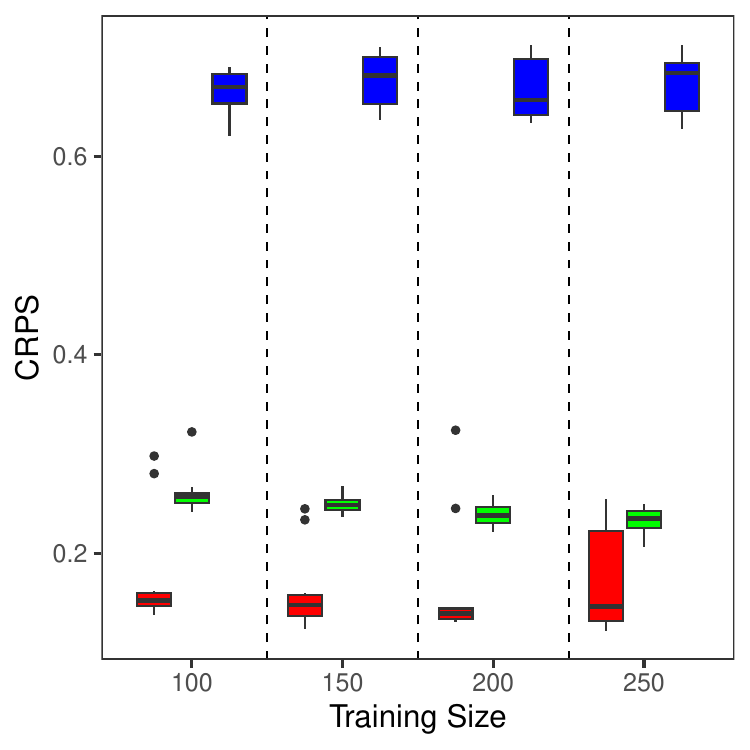}
\caption{Results for the modified logarithmic function example with input gaps.  Boxplots summarize out-of-sample performance with circular RMSE (left) and CRPS (right) for all three WGP methods across $10$ random training sets of increasing size.  Smaller values indicate higher accuracy.}
\label{fig:log_gap_results}
\end{figure}
Empirical performance results are provided in Figure \ref{fig:log_gap_results}.  While our method exhibits more variability in performance than before, it achieves the lowest circular RMSE across the three methods.  Likewise, our WGP achieves the smallest CRPS values.

\section{RFID Tag Localization} \label{sec:results}
In this section we return to the motivating RFID tag localization problem introduced in Section \ref{sec:intro}.  Data collected from experiments in a mock RFID lab setting are discussed in Section \ref{ss:rfid_exp}.  Our WGP-based hierarchical modeling approach for the purposes of RFID tag localization is outlined in Section \ref{ss:rfid_hier}.

\subsection{RFID Localization} \label{ss:rfid}
Passive RFID tags are commonly used to keep track of inventory in indoor environments.  Tags with unique Electronic Product Codes (EPCs) are attached to items; a fixed antenna emits signals that propagate through the environment, and signals that reflect off these tags are processed by a reader which identifies these EPCs.  While helpful for identifying the existence of particular assets, passive RFID technology provides no direct information regarding location.  This information would significantly expedite inventory tracking, especially in environments with many tags.

It has therefore been of interest to researchers to use available RFID information to infer tag location indirectly \citep{bouet2008rfid}.  A number of measurements have been explored as informative markers of tag position.  One popular metric is the Received Signal Strength Indicator (RSSI), which measures the strength of the signal received by the reader.  However, RSSI has been found to be an unreliable measure for localization in indoor environments due to the prevalence of back-scattering, in which signals propagate off walls and deteriorate signal strength \citep{chandrasekaran2009empirical}.  
\begin{figure}[ht!]
    \centering
    \includegraphics[width=0.8\linewidth]{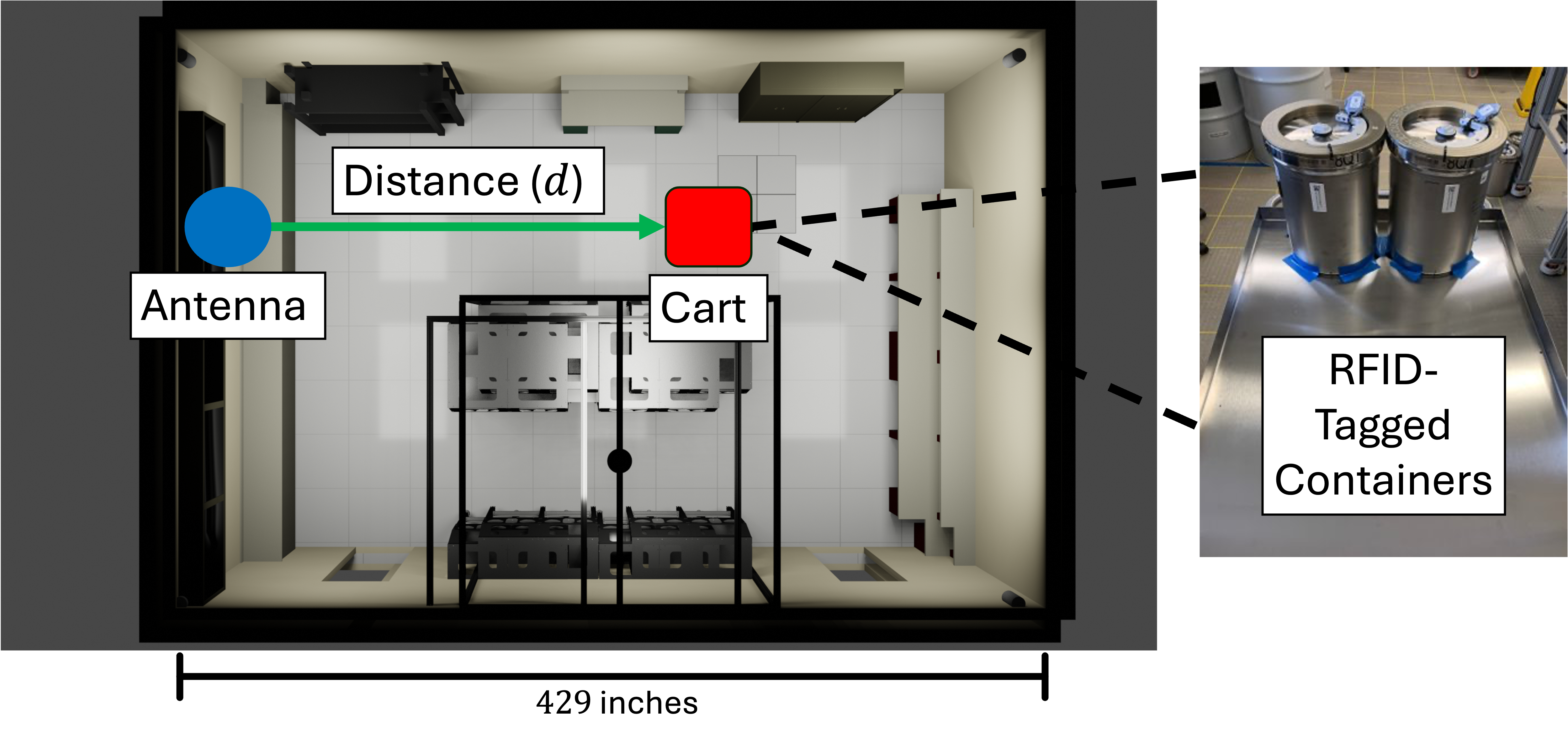}
    \caption{Top-down view of three-dimensional rendering of lab environment.}
    \label{fig:lab}
\end{figure}
This issue is compounded in laboratory settings such as the one shown in Figure \ref{fig:lab}; metallic surfaces on containers (shown in the picture on the right) and gloveboxes installed to safely handle materials (shown from a top-down perspective by the silver rectangles) are conducive to signal scattering.  This combined with a cluttered environment impedes signal strength and leads to noisier RSSI.

Another measure used for localization is the ``phase shift'', also known as the phase angle, which describes the change in a waveform's position after communicating with an RFID tag.  
\begin{figure}[ht!]
\centering
\includegraphics[width=0.32\linewidth,trim=20 40 0 30]{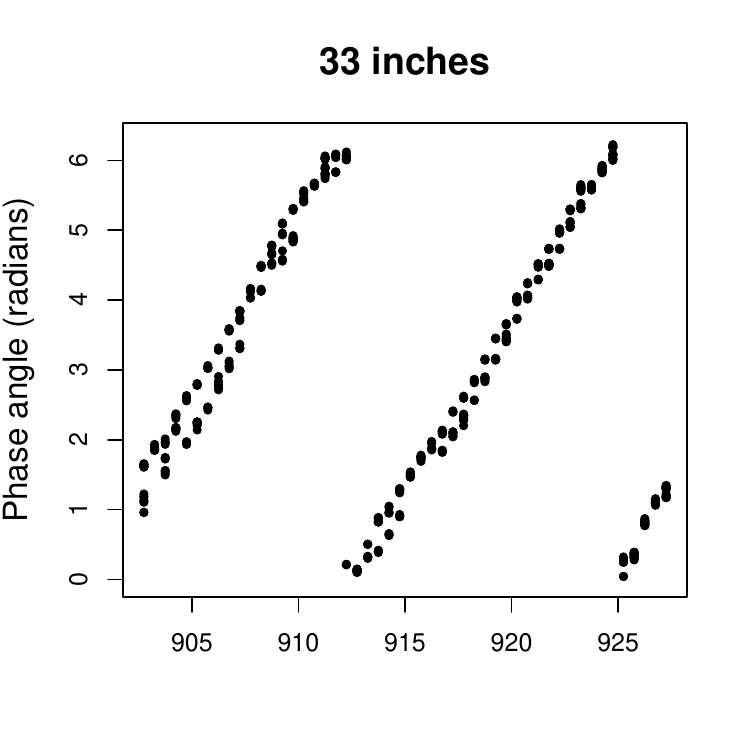}
\includegraphics[width=0.32\linewidth,trim=20 40 0 30]{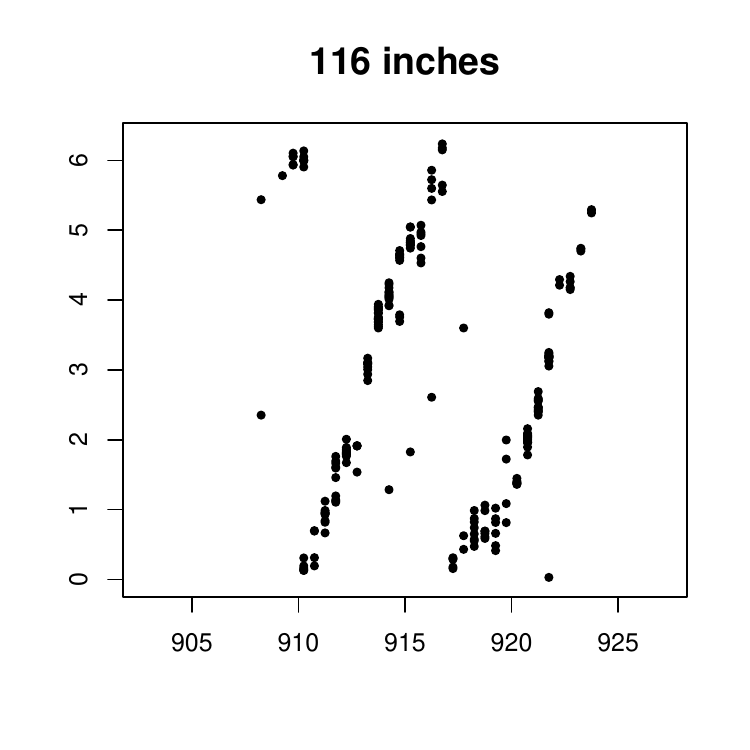}
\includegraphics[width=0.32\linewidth,trim=20 40 0 30]{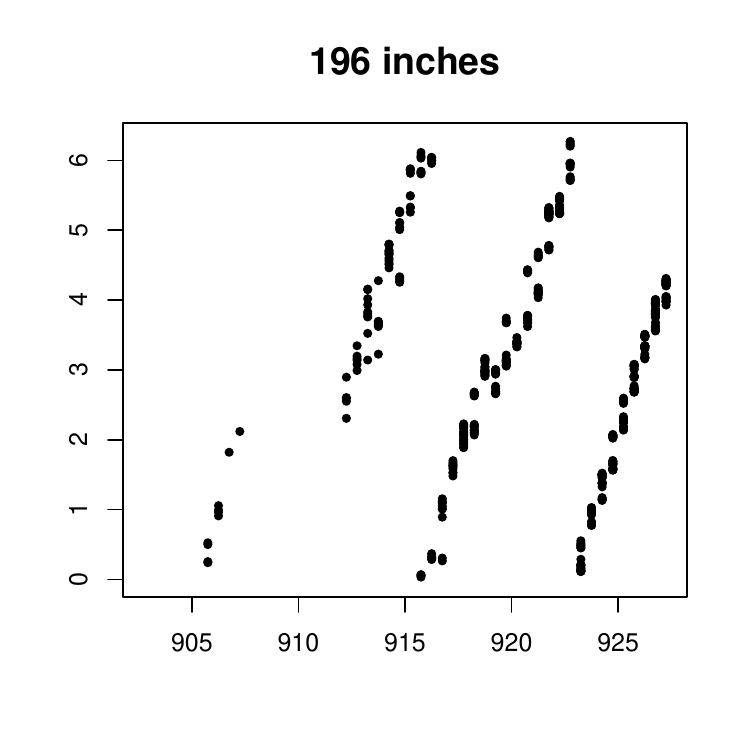}
\includegraphics[width=0.32\linewidth,trim=20 0 0 40]{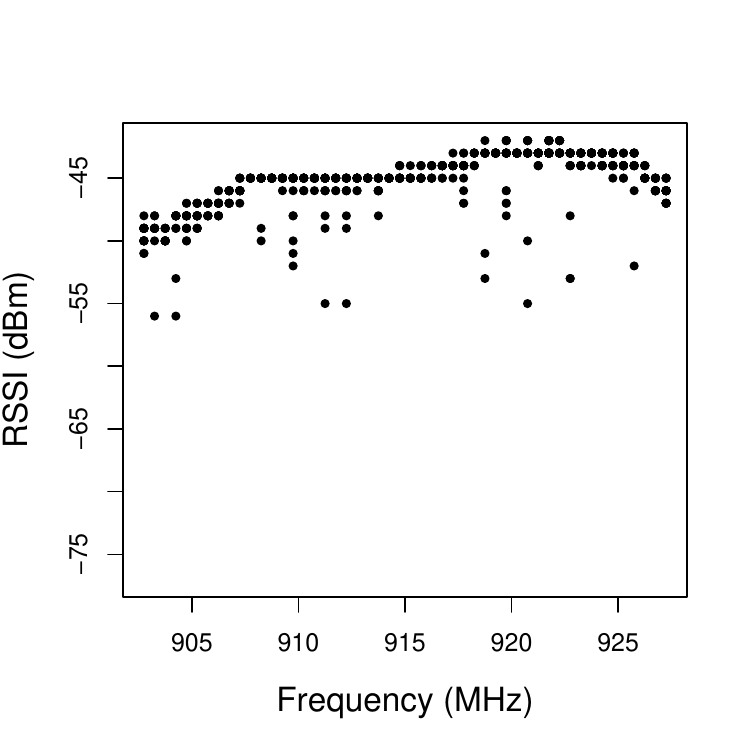}
\includegraphics[width=0.32\linewidth,trim=20 0 0 40]{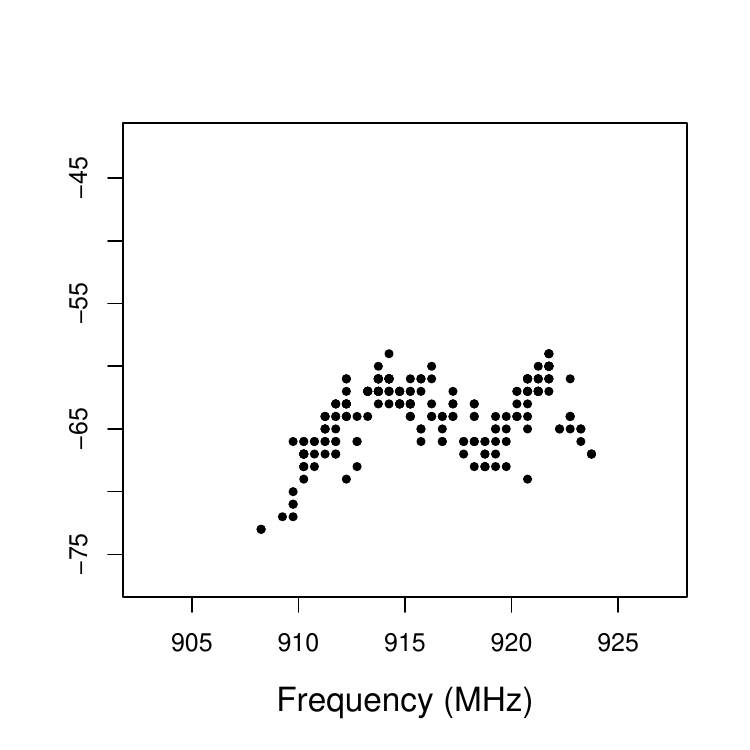}
\includegraphics[width=0.32\linewidth,trim=20 0 0 40]{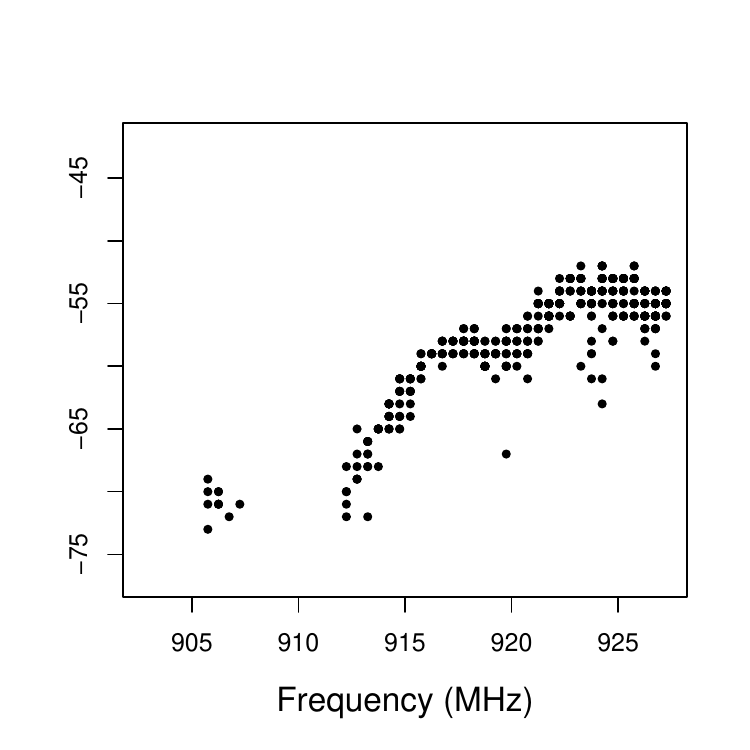}
\caption{Phase angles and RSSI values observed at different frequencies for increasing distance.}
\label{fig:rfid_eda}
\end{figure}
While phase angle has been shown to be a more reliable metric in indoor environments compared to RSSI \citep{liu2014anchor}, it also suffers from noise and missing data in our laboratory setting, as can be seen by the plots in Figure \ref{fig:rfid_eda}.  Signals below an RSSI threshold of around $-75$ dBm fail to be properly read, leading to gaps of frequencies for which phase angle is missing.  What's more, weaker signals typically observed at farther distances tend to produce noisier phase angles, including outliers like those seen in the top-middle plot.

Recent research using phase angle for RFID localization hinges on the theoretical relationship between signal frequency $f$, phase angle $\phi$, and distance to tag $d$:
\begin{equation}
    d = \dfrac{c}{4\pi}\dfrac{\partial \phi}{\partial f},
\label{eq:phase_freq}
\end{equation}
where $c$ is the speed of light constant \citep{shangguan2015relative}.  Eq.~(\ref{eq:phase_freq}) presumes distance is proportional to the derivative of phase with respect to frequency.  That is, inference of the rate of change of phase angle with respect to frequency provides insight into how far away the tag is.  Our goal is then to appropriately estimate $\beta \equiv \frac{\partial\phi}{\partial f}$ for the purpose of inferring tag distance.

\subsection{RFID Experiments} \label{ss:rfid_exp}
Data were collected from experiments in a mock lab setting shown in Figure \ref{fig:lab}.  Two RFID-tagged containers were placed on a metallic cart (shown on the right) and moved along a section of the lab in a straight line.  A fixed antenna was placed on a desk at the end of the lab and aligned with the cart path.  The cart was placed at $22$ unique distances generated from a Latin hypercube sampling (LHS) design ranging between $0$ and about $429$ inches (the maximal length allowable by the dimensions of the lab).  At each distance, an RFID scan was performed for two minutes. This scan automatically cycles through $50$ unique frequency channels ranging from $902.75$ to $927.25$ MHz in an order dictated by a fixed frequency hopping table.

Phase-frequency data collected by one of the tags facing the antenna for three example tests are plotted in the top row of Figure \ref{fig:rfid_eda}.  While they all exhibit a general positive linear trend, the slope of the line for the largest distance (right) appears to be larger in magnitude than for the smallest distance (left), leading to more apparent wrapping.  We also tend to observe noisier phase angle behavior and more frequent outliers as the distance from antenna to cart increases.

\subsection{Hierarchical Model} \label{ss:rfid_hier}
Based on Eq.~(\ref{eq:phase_freq}), we have reason to believe distance is informative of each test's phase-frequency relationship.  Rather than a fit separate WGP model for each test independent of one another, we wish to share information across tests in estimating their respective slope values.  More specifically, slope values for different tests should be more similar to each other if their distances are close, and vice versa.  Capturing this three-way relationship is also crucial for solving the inverse problem; given the distance for a given test is unobserved, how can we use the estimated phase-frequency relationship to ``impute'' it?

To achieve this, we propose a hierarchical approach to modeling phase angle.
\begin{figure}[ht!]
    \centering
    \hspace{-10mm}
    \includegraphics[width=0.8\linewidth]{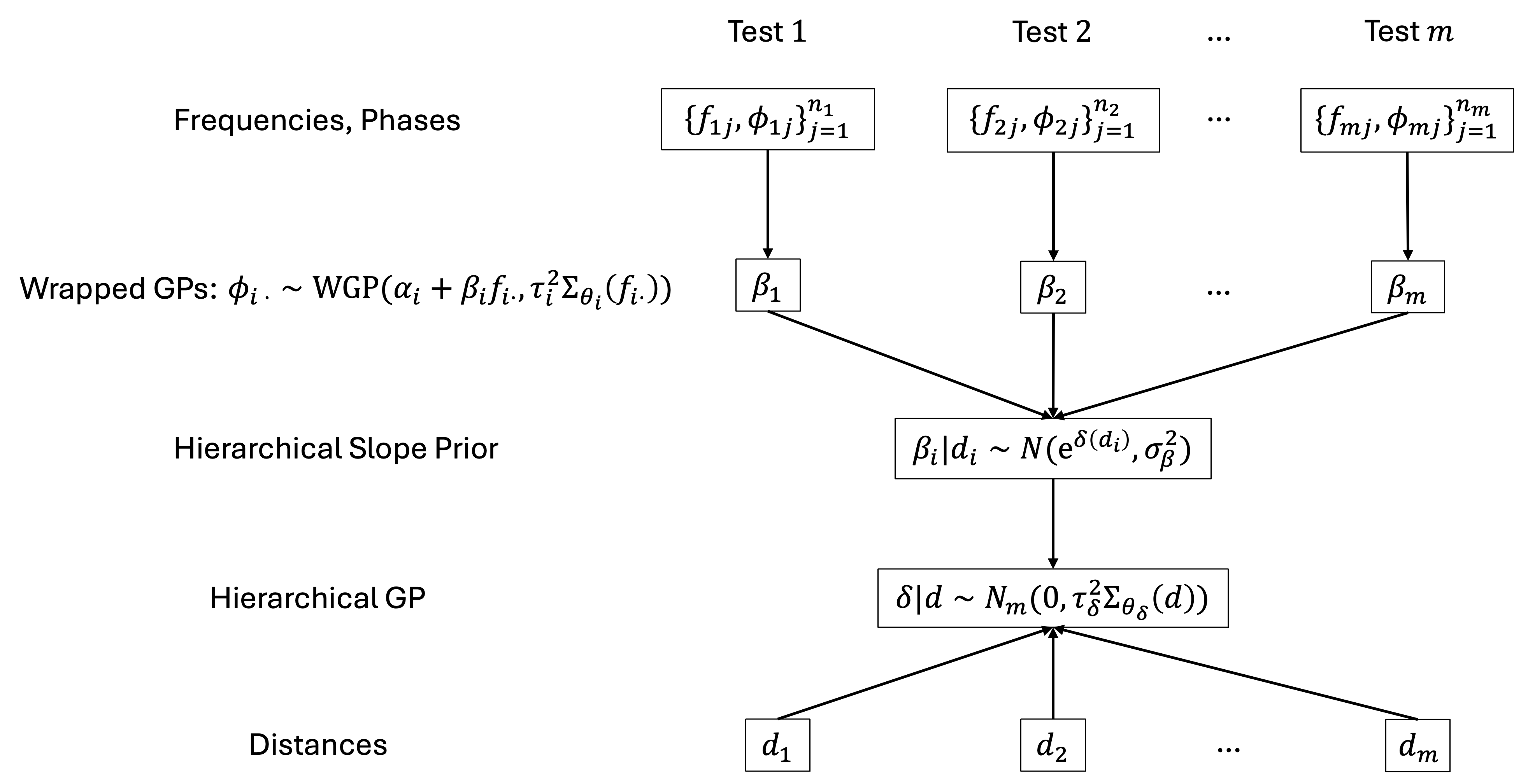}
    \caption{Hierarchical modeling approach for RFID experiments.}
    \label{fig:hier}
\end{figure}
Our model is illustrated in Figure \ref{fig:hier}; at the lowest level, we have individual phase-frequency pairs $\{f_{ij},\phi_{ij}\}_{j=1}^{n_j}$ for tests $i = 1,\dots,m$.  We use our WGP model outlined in Section \ref{sec:wgp} to model phase angle as a function of frequency, which we express using the shorthand \\$\phi_{i\cdot} \sim \text{WGP}\left(\alpha_i + \beta_if_{i \cdot}, \tau_i^2\Sigma_{\theta_i}(f_{i\cdot})\right)$, with a modification to the prior for $\beta_i$.  For each test we specify a hierarchical prior $\beta_i \mid d_i \sim \mathcal{N}\left(e^{\delta(d_i)}, \sigma_\beta^2\right)$, where $d_i$ is the distance for the $i^\text{th}$ test.  The vector $\delta\in\mathbb{R}^m$ corresponds to the output of an additional latent hierarchical mean GP, where $\delta \mid d \sim \mathcal{N}_m\left(0, \tau_\delta^2\Sigma_{\theta_\delta}(d)\right)$.  Since $\delta$ uses distance as an input, we use information of $d_i$ to inform the prior mean for $\beta_i$;  that is, a priori we assume the correlation between slopes of different tests are inversely proportional to their relative distances.  By exponentiating $\delta(d_i)$ we ensure each prior mean remains positive.  We construct the $\delta$-level covariance via the squared-exponential kernel as we do at the WGP-level by replacing $x$ with $d$ in Eq.~(\ref{eq:sq_exp}).  We specify a smaller common slope variance parameter $\sigma_\beta^2 = 0.1$ so the prior density for each test is more heavily concentrated around the distance-informed mean.

\begin{algorithm}[ht!]
\DontPrintSemicolon
{\bf Input:} Starting values for WGP-level parameters for tests $i=1,\dots,m$, starting values for hierarchical parameters $\delta^{(1)}, \theta_\delta^{(1)},\tau_\delta^{2(1)}$\\
{\bf Output:} $T$ posterior samples from WGP-level parameters and hierarchical parameters $\delta,\theta_\delta$\\
~
\For{$t \in 1,\dots,T$}{
\For{$i \in 1,\dots,m$}{
Construct hierarchical prior $\beta_i \sim \mathcal{N}\left(e^{\delta^{(t)}_i}, \sigma_\beta^2\right)$\\
Draw parameters corresponding to $\text{WGP}\left(\alpha_i + \beta_if_{i \cdot}, \Sigma_{\theta_i}(f_{i\cdot})\right)$ \tcp*{Alg. (\ref{alg:mcmc})}
}
Draw $\tau_\delta^{2(t+1)} \sim \tau_\delta^2 \mid \theta_\delta^{(t)},\delta^{(t)},d$ \tcp*{Gibbs}
Draw $\theta_\delta^{(t+1)} \sim \theta_\delta \mid \theta_\delta^{(t)},\tau_\delta^{2(t+1)},\delta^{(t)},d$ \tcp*{MH}
Draw $\delta^{(t+1)} \sim \delta \mid \delta^{(t)},\theta_\delta^{(t+1)},\tau_\delta^{2(t+1)},\beta_1^{(t+1)},\dots,\beta_m^{(t+1)},d$ \tcp*{ESS}
}
\caption{MCMC procedure for hierarchical WGP estimation}
\label{alg:hier}
\end{algorithm}

We estimate the individual WGP models as well as hierarchical parameters corresponding to $\delta$ in a single MCMC scheme, detailed in Algorithm \ref{alg:hier}.  Those corresponding to individual test-level WGP parameters are sampled from their respective full conditionals at each MCMC iteration, utilizing a prior mean of $e^{\delta_i^{(t)}}$ for the $i^\text{th}$ slope parameter.  Scales and lengthscales are sampled from their posteriors similar to how they are at the WGP-level, only now replacing $X$ with $d$ and $Y$ with $\delta^{(t)}$.  

Since $\delta$ has a Gaussian prior, we can once more use ESS to generate new posterior samples.  Similar in spirit to Algorithm \ref{alg:ess}, the only substantial change is in the likelihood used to evaluate proposals, which at this layer of the model now corresponds to the test-level slopes $\beta_1,\dots,\beta_m$.  As the slopes are conditionally pairwise independent given $\delta$, we can express the hierarchical log-likelihood as 
\begin{equation}
\ell(\delta\mid \beta_1,\dots,\beta_m) = \sum_{i=1}^m \ell(\delta_i\mid\beta_i) = \sum_{i=1}^m \log\left(\phi\left(\dfrac{\beta_i - e^{\delta_i}}{\sigma_\beta}\right)\right),
\label{eq:log_like}
\end{equation}
where $\sigma_\beta^2 = 0.1$.  In Algorithm \ref{alg:hier} we use the samples $\beta_1^{(t)},\dots,\beta_m^{(t)}$ accepted at the $t^\text{th}$ iteration to evaluate $\delta^{(t)}$ and $\delta^*$.

\vspace{5mm}
\begin{figure}[ht!]
\centering
\includegraphics[width=0.32\linewidth,trim=20 10 0 50]{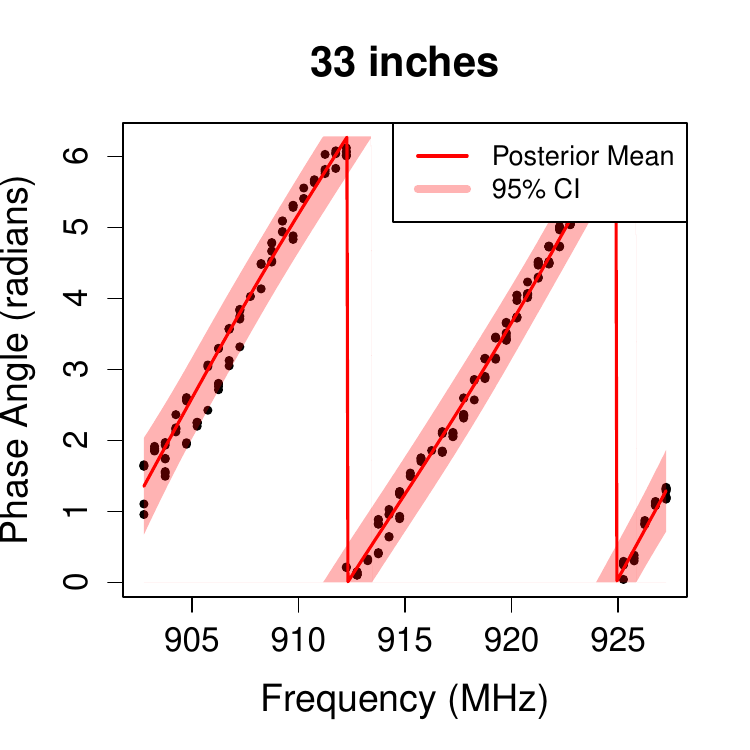}
\includegraphics[width=0.32\linewidth,trim=20 10 0 50]{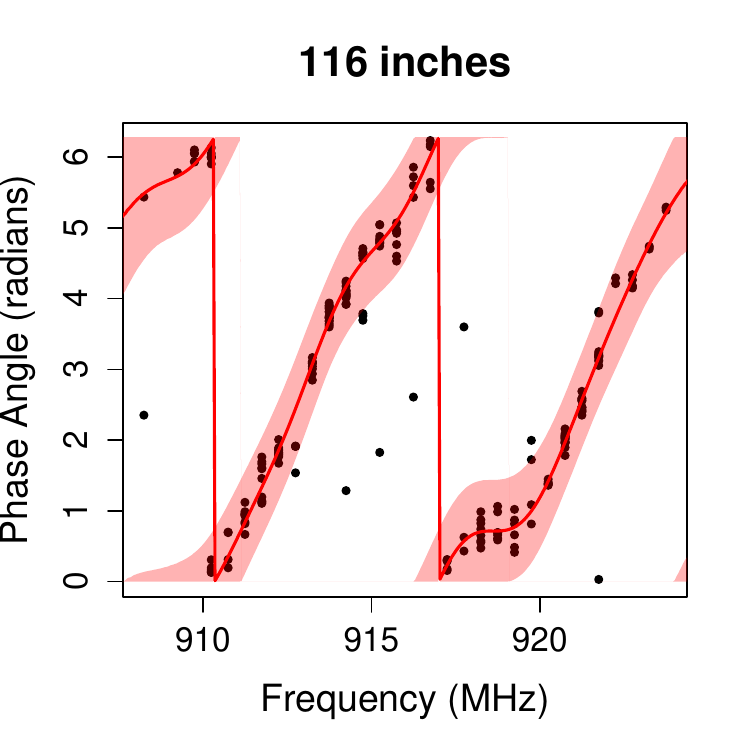}
\includegraphics[width=0.32\linewidth,trim=20 10 0 50]{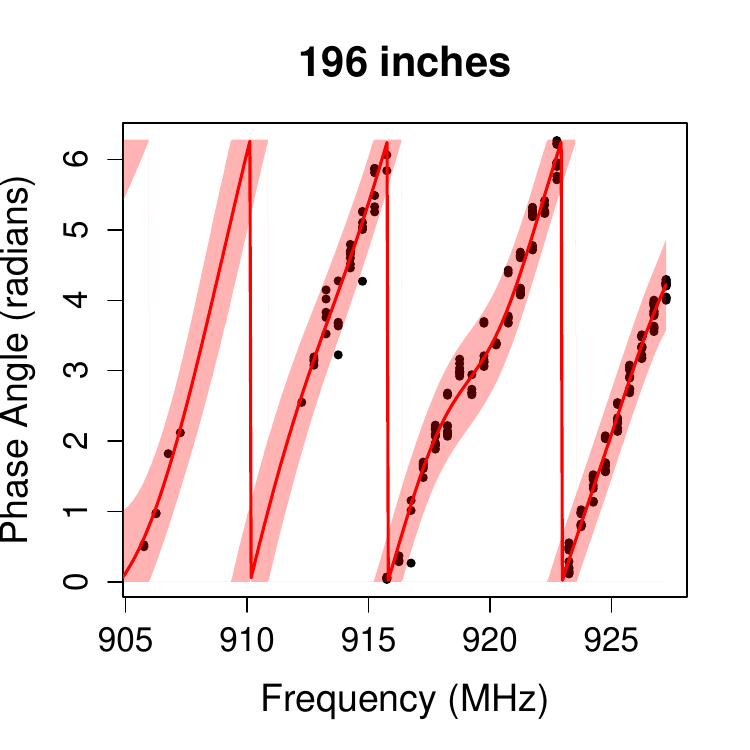}
\caption{Posterior mean and $95\%$ credible interval predictions from our WGP model for three example RFID tests.}
\label{fig:phase_fit}
\end{figure}
Test-level WGP fits produced by this model for example tests are visualized in Figure \ref{fig:phase_fit}.  Our WGP approach correctly identifies the wrapping behavior despite common outliers observed.  What's more, it can identify what appears to be a wrapping in the interval of missing frequencies in the right panel.

\begin{figure}[ht!]
\centering
\includegraphics[scale=0.55,trim=20 20 0 20]{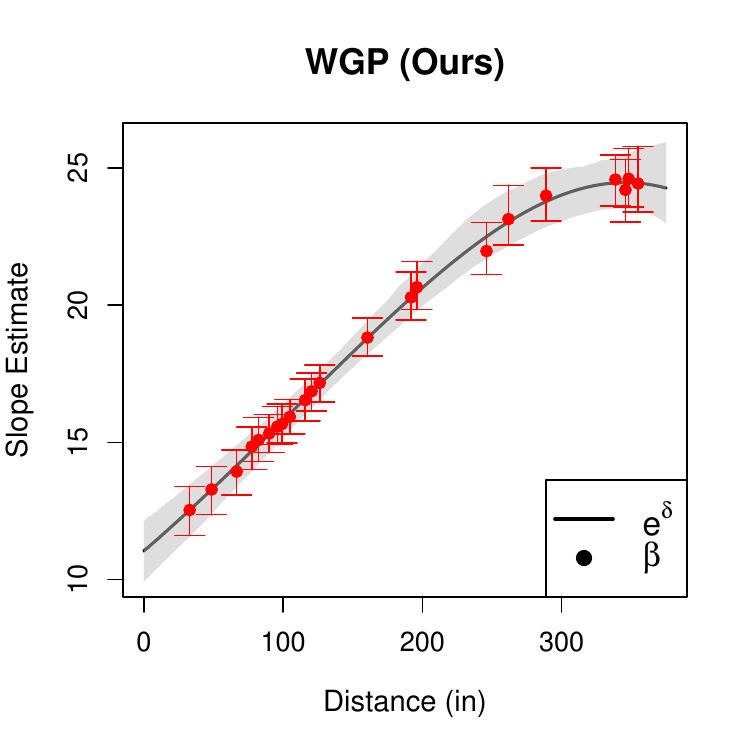}
\includegraphics[scale=0.55,trim=20 20 0 20]{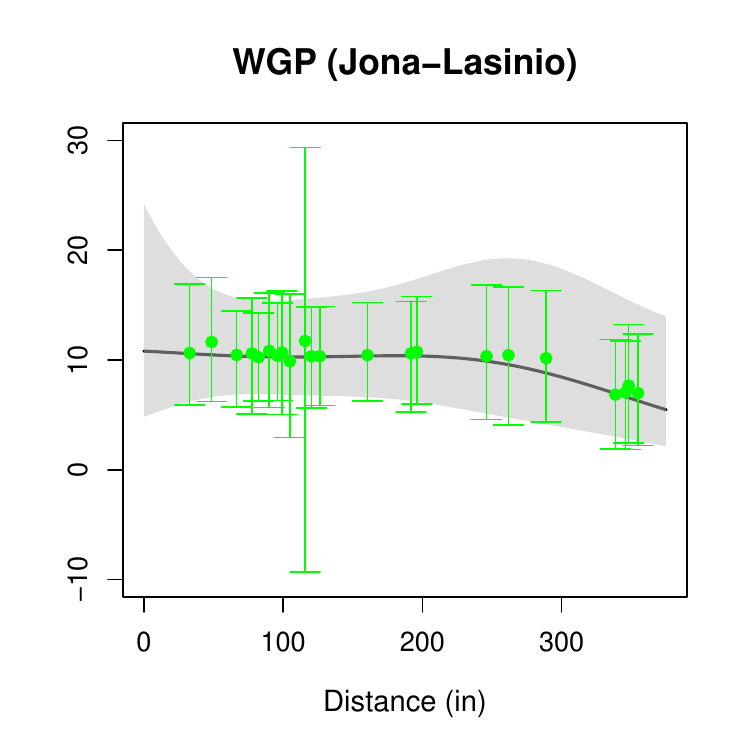}
\caption{Comparison of slope and hierarchical mean GP estimation.  Points/error bars denote slope estimation, while black lines and gray intervals show the exponent of the hierarchical mean GP.}
\label{fig:phase_slope}
\end{figure}
Figure \ref{fig:phase_slope} visually compares estimates of the phase-frequency relationship for all $22$ tests as a function of distance between our proposed WGP and that of \cite{jona2012spatial}.  The points and error bars represent posterior mean and $95\%$ quantile intervals, respectively.   Black lines and gray intervals show mean and $95\%$ credible region estimates of the (exponentiated) hierarchical mean GP $\delta$.  Our method better captures the known increasing trend in slope.  Along with a general (but seemingly non-linear) positive trend in slope estimation for increasing distance, we observe larger slope uncertainty for larger distances than we do for smaller ones.  This likely stems from less (and typically noisier) data observed at these distances.

\section{Discussion} \label{sec:discuss}
We have proposed a new approach to angular regression that is tailored to a unique class of problems.  By taking advantage of monotonic behavior in $K$, we have improved estimation accuracy of the wrapped GP model.  We have also made use of a remarkably effective sampling scheme (ESS) to estimate our model with full UQ, allowing for more robust inference in settings with noisy responses by assuming Student's $t$ errors.  This has been evident in the superior empirical performance on simulated examples compared to existing wrapped GP methods.  While the RFID problem discussed in Section \ref{sec:results} is the only real-world discussed in this paper, we believe there are other settings where monotonic wrapping behavior is observed.  In the study of fluid dynamics, the direction of vortices as a function of time may exhibit this structure, for instance \citep{rogers1987structure}.

While the concept of monotonic wrapping behavior is apparent for one-dimensional $X$, its extension to higher dimension is not clear.  Monotonicity when there two or more inputs is not well-defined since there is no natural ordering such that $K$ is increasing in every direction.  Extension of our proposed methods to higher dimension is therefore non-trivial.  However, generalizing our work to non-monotonic wrapping behavior may be viable with extensions to the proposals for $K$ outlined in Section \ref{ss:wrap_num}.  Our framework for estimating the wrapping locations and partitioning the input space could potentially allow for wrapping in both directions.  This would require changes to new $K$ proposals, or the introduction of new adjustments to the partitioning that allow for non-monotonicity (growing a wrapping number in the opposite direction, for instance).  While this would open our method to a wider class of angular regression settings, we believe the assumption of monotonic wrapping significantly contributes to our model's performance.  

The model proposed in Section \ref{ss:rfid_hier} shows promise in capturing the phase-frequency-distance relationship, but the task of RFID tag localization remains an open problem.  Solving the inverse problem of inferring distance based on frequency and phase is challenging in this setting since we rely on distance as input for the hierarchical GP $\delta$.  Imputation of distance for new tests may lead to identifiability issues, and may require more unique distances to properly estimate.  Furthermore, the need to estimate parameters for tests with both observed and missing distances simultaneously makes scaling difficult.  Approximations might be able to allow for out-of-sample prediction for new tests without the need for re-fitting our model from scratch.

\section*{Acknowledgement}
We acknowledge LANL staff Katie Bridgewater and Rollin Lakis for general project management; Allison Davis, Andre Green, and Alexander Miller for software development and assistance in performing RFID experiments.

\section*{Funding}
Our work is funded by the U.S. National Nuclear Security Agency, NA-191, under the Dynamic Material Control (DYMAC) collaboration.

\bibliography{manuscript}

@book{mardia2000directional,
  title={Directional statistics},
  author={Mardia, Kanti V and Jupp, Peter E},
  year={2000},
  publisher={John Wiley \& Sons}
}

@article{jones2005family,
  title={A family of symmetric distributions on the circle},
  author={Jones, MC and Pewsey, Arthur},
  journal={Journal of the American Statistical Association},
  volume={100},
  number={472},
  pages={1422--1428},
  year={2005},
  publisher={Taylor \& Francis}
}

@article{jona2012spatial,
  title={Spatial analysis of wave direction data using wrapped Gaussian processes},
  author={Jona-Lasinio, Giovanna and Gelfand, Alan and Jona-Lasinio, Mattia},
journal={The Annals of Applied Statistics},
  year={2012}
}

@inproceedings{mallasto2018wrapped,
  title={Wrapped Gaussian process regression on Riemannian manifolds},
  author={Mallasto, Anton and Feragen, Aasa},
  booktitle={Proceedings of the IEEE Conference on Computer Vision and Pattern Recognition},
  pages={5580--5588},
  year={2018}
}

@book{jammalamadaka2001topics,
  title={Topics in circular statistics},
  author={Jammalamadaka, S Rao and Sengupta, Ambar},
  volume={5},
  year={2001},
  publisher={world scientific}
}

@article{coles1998inference,
  title={Inference for circular distributions and processes},
  author={Coles, Stuart},
  journal={Statistics and Computing},
  volume={8},
  pages={105--113},
  year={1998},
  publisher={Springer}
}

@article{ravindran2011bayesian,
  title={Bayesian analysis of circular data using wrapped distributions},
  author={Ravindran, Palanikumar and Ghosh, Sujit K},
  journal={Journal of Statistical Theory and Practice},
  volume={5},
  number={4},
  pages={547--561},
  year={2011},
  publisher={Taylor \& Francis}
}

@article{pigoli2016kriging,
  title={Kriging prediction for manifold-valued random fields},
  author={Pigoli, Davide and Menafoglio, Alessandra and Secchi, Piercesare},
  journal={Journal of Multivariate Analysis},
  volume={145},
  pages={117--131},
  year={2016},
  publisher={Elsevier}
}

@article{fisher1992regression,
  title={Regression models for an angular response},
  author={Fisher, Nicholas I and Lee, Alan J},
  journal={Biometrics},
  pages={665--677},
  year={1992},
  publisher={JSTOR}
}

@inproceedings{liu2014anchor,
  title={Anchor-free backscatter positioning for RFID tags with high accuracy},
  author={Liu, Tianci and Yang, Lei and Lin, Qiongzheng and Guo, Yi and Liu, Yunhao},
  booktitle={IEEE INFOCOM 2014-IEEE Conference on Computer Communications},
  pages={379--387},
  year={2014},
  organization={IEEE}
}

@inproceedings{chandrasekaran2009empirical,
  title={Empirical evaluation of the limits on localization using signal strength},
  author={Chandrasekaran, Gayathri and Ergin, Mesut Ali and Yang, Jie and Liu, Song and Chen, Yingying and Gruteser, Marco and Martin, Richard P},
  booktitle={2009 6th Annual IEEE Communications Society Conference on Sensor, Mesh and Ad Hoc Communications and Networks},
  pages={1--9},
  year={2009},
  organization={IEEE}
}

@inproceedings{bouet2008rfid,
  title={RFID tags: Positioning principles and localization techniques},
  author={Bouet, Mathieu and Dos Santos, Aldri L},
  booktitle={2008 1st IFIP Wireless Days},
  pages={1--5},
  year={2008},
  organization={Ieee}
}

@inproceedings{shangguan2015relative,
  title={Relative localization of $\{$RFID$\}$ tags using $\{$Spatial-Temporal$\}$ phase profiling},
  author={Shangguan, Longfei and Yang, Zheng and Liu, Alex X and Zhou, Zimu and Liu, Yunhao},
  booktitle={12th USENIX Symposium on networked systems design and implementation (NSDI 15)},
  pages={251--263},
  year={2015}
}

@article{yang2021rfid,
  title={RFID tag localization with a sparse tag array},
  author={Yang, Chao and Wang, Xuyu and Mao, Shiwen},
  journal={IEEE internet of things journal},
  volume={9},
  number={18},
  pages={16976--16989},
  year={2021},
  publisher={IEEE}
}

@inproceedings{kurz2014efficient,
  title={Efficient evaluation of the probability density function of a wrapped normal distribution},
  author={Kurz, Gerhard and Gilitschenski, Igor and Hanebeck, Uwe D},
  booktitle={2014 Sensor Data Fusion: Trends, Solutions, Applications (SDF)},
  pages={1--5},
  year={2014},
  organization={IEEE}
}

@article{ferrari2009wrapping,
  title={The wrapping approach for circular data Bayesian modeling},
  journal={Ph.D Dissertation},
  author={Ferrari, Clarissa},
  year={2009},
  publisher={alma}
}

@article{bell2024review,
  title={A Review of Wrapped Distributions for Circular Data},
  author={Bell, William and Nadarajah, Saralees},
  journal={Mathematics},
  volume={12},
  number={16},
  pages={2440},
  year={2024},
  publisher={MDPI}
}

@article{lange1989robust,
  title={Robust statistical modeling using the t distribution},
  author={Lange, Kenneth L and Little, Roderick JA and Taylor, Jeremy MG},
  journal={Journal of the American Statistical Association},
  volume={84},
  number={408},
  pages={881--896},
  year={1989},
  publisher={Taylor \& Francis}
}

@inproceedings{murray2010elliptical,
  title={Elliptical slice sampling},
  author={Murray, Iain and Adams, Ryan and MacKay, David},
  booktitle={Proceedings of the thirteenth international conference on artificial intelligence and statistics},
  pages={541--548},
  year={2010},
  organization={JMLR Workshop and Conference Proceedings}
}

@article{barnett2025monotonic,
  title={Monotonic warpings for additive and deep Gaussian processes},
  author={Barnett, Steven D and Beesley, Lauren J and Booth, Annie S and Gramacy, Robert B and Osthus, Dave},
  journal={Statistics and Computing},
  volume={35},
  number={3},
  pages={65},
  year={2025},
  publisher={Springer}
}

@article{sauer2023active,
  title={Active learning for deep Gaussian process surrogates},
  author={Sauer, Annie and Gramacy, Robert B and Higdon, David},
  journal={Technometrics},
  volume={65},
  number={1},
  pages={4--18},
  year={2023},
  publisher={Taylor \& Francis}
}

@article{cressie1988spatial,
  title={Spatial prediction and ordinary kriging},
  author={Cressie, Noel},
  journal={Mathematical geology},
  volume={20},
  number={4},
  pages={405--421},
  year={1988},
  publisher={Springer}
}

@article{chipman2010bart,
  title={BART: Bayesian additive regression trees},
  author={Chipman, Hugh A and George, Edward I and McCulloch, Robert E},
  journal={Annals of applied statistics},
  year={2010}
}

@article{carta2008joint,
  title={A joint probability density function of wind speed and direction for wind energy analysis},
  author={Carta, Jos{\'e} A and Ramirez, Penelope and Bueno, Celia},
  journal={Energy Conversion and Management},
  volume={49},
  number={6},
  pages={1309--1320},
  year={2008},
  publisher={Elsevier}
}

@book{breckling2012analysis,
  title={The analysis of directional time series: applications to wind speed and direction},
  author={Breckling, Jens},
  volume={61},
  year={2012},
  publisher={Springer Science \& Business Media}
}

@inproceedings{mcgraw2006mises,
  title={Von Mises-Fisher mixture model of the diffusion ODF},
  author={McGraw, Tim and Vemuri, Baba C and Yezierski, Bob and Mareci, Thomas},
  booktitle={3rd IEEE International Symposium on Biomedical Imaging: Nano to Macro, 2006.},
  pages={65--68},
  year={2006},
  organization={IEEE}
}

@article{ryali2013parcellation,
  title={A parcellation scheme based on von Mises-Fisher distributions and Markov random fields for segmenting brain regions using resting-state fMRI},
  author={Ryali, Srikanth and Chen, Tianwen and Supekar, Kaustubh and Menon, Vinod},
  journal={Neuroimage},
  volume={65},
  pages={83--96},
  year={2013},
  publisher={Elsevier}
}

@article{janaswamy2002angle,
  title={Angle of arrival statistics for a 3-D spheroid model},
  author={Janaswamy, Ramakrishna},
  journal={IEEE Transactions on vehicular technology},
  volume={51},
  number={5},
  pages={1242--1247},
  year={2002},
  publisher={IEEE}
}

@inproceedings{nikitin2010phase,
  title={Phase based spatial identification of UHF RFID tags},
  author={Nikitin, Pavel V and Martinez, Rene and Ramamurthy, Shashi and Leland, Hunter and Spiess, Gary and Rao, KVS},
  booktitle={2010 IEEE International Conference on RFID (IEEE RFID 2010)},
  pages={102--109},
  year={2010},
  organization={IEEE}
}

@article{rivest2016general,
  title={A general angular regression model for the analysis of data on animal movement in ecology},
  author={Rivest, Louis-Paul and Duchesne, Thierry and Nicosia, Aur{\'e}lien and Fortin, Daniel},
  journal={Journal of the Royal Statistical Society Series C: Applied Statistics},
  volume={65},
  number={3},
  pages={445--463},
  year={2016},
  publisher={Oxford University Press}
}

@book{coleman2017analysis,
  title={Analysis and modeling of radio wave propagation},
  author={Coleman, Christopher John},
  year={2017},
  publisher={Cambridge University Press}
}

@article{gao2006application,
  title={On the application of the von Mises distribution and angular regression methods to investigate the seasonality of disease onset},
  author={Gao, Fei and Chia, Kee-Seng and Krantz, Ingela and Nordin, Per and Machin, David},
  journal={Statistics in medicine},
  volume={25},
  number={9},
  pages={1593--1618},
  year={2006},
  publisher={Wiley Online Library}
}

@Inbook{laha2022,
author="Laha, Arnab Kumar
and Majumdar, Sourav",
title="Angular-Angular and Linear-Angular Regression Using ANN",
bookTitle="Directional Statistics for Innovative Applications: A Bicentennial Tribute to Florence Nightingale",
year="2022",
publisher="Springer Nature Singapore",
address="Singapore",
pages="451--474",
isbn="978-981-19-1044-9",
doi="10.1007/978-981-19-1044-9_24",
url="https://doi.org/10.1007/978-981-19-1044-9_24"
}

@article{jammalamadaka2004new,
  title={New families of wrapped distributions for modeling skew circular data},
  author={Jammalamadaka, Sreenivasa Rao and Kozubowski, Tomasz J},
  journal={Communications in Statistics-Theory and Methods},
  volume={33},
  number={9},
  pages={2059--2074},
  year={2004},
  publisher={Taylor \& Francis}
}

@article{downs2002circular,
  title={Circular regression},
  author={Downs, Thomas D and Mardia, KV},
  journal={Biometrika},
  volume={89},
  number={3},
  pages={683--698},
  year={2002},
  publisher={Oxford University Press}
}

@book{williams2006gaussian,
  title={Gaussian processes for machine learning},
  author={Williams, Christopher KI and Rasmussen, Carl Edward},
  volume={2},
  year={2006},
  publisher={MIT press Cambridge, MA}
}

@article{tran2015variational,
  title={The variational Gaussian process},
  author={Tran, Dustin and Ranganath, Rajesh and Blei, David M},
  journal={arXiv preprint arXiv:1511.06499},
  year={2015}
}

@article{mardia1975statistics,
  title={Statistics of directional data},
  author={Mardia, Kantilal Varichand},
  journal={Journal of the Royal Statistical Society Series B: Statistical Methodology},
  volume={37},
  number={3},
  pages={349--371},
  year={1975},
  publisher={Oxford University Press}
}

@article{geomstats,
  author  = {Nina Miolane and Nicolas Guigui and Alice Le Brigant and Johan Mathe and Benjamin Hou and Yann Thanwerdas and Stefan Heyder and Olivier Peltre and Niklas Koep and Hadi Zaatiti and Hatem Hajri and Yann Cabanes and Thomas Gerald and Paul Chauchat and Christian Shewmake and Daniel Brooks and Bernhard Kainz and Claire Donnat and Susan Holmes and Xavier Pennec},
  title   = {Geomstats:  A Python Package for Riemannian Geometry in Machine Learning},
  journal = {Journal of Machine Learning Research},
  year    = {2020},
  volume  = {21},
  number  = {223},
  pages   = {1-9},
  url     = {http://jmlr.org/papers/v21/19-027.html}
}

@article{holsclaw2013gaussian,
  title={Gaussian process modeling of derivative curves},
  author={Holsclaw, Tracy and Sans{\'o}, Bruno and Lee, Herbert KH and Heitmann, Katrin and Habib, Salman and Higdon, David and Alam, Ujjaini},
  journal={Technometrics},
  volume={55},
  number={1},
  pages={57--67},
  year={2013},
  publisher={Taylor \& Francis}
}

@article{fisher1994time,
  title={Time series analysis of circular data},
  author={Fisher, NI and Lee, AJ12819370796},
  journal={Journal of the Royal Statistical Society: Series B (Methodological)},
  volume={56},
  number={2},
  pages={327--339},
  year={1994},
  publisher={Wiley Online Library}
}

@inproceedings{fletcher2011geodesic,
  title={Geodesic regression on Riemannian manifolds},
  author={Fletcher, Thomas},
  booktitle={Proceedings of the Third International Workshop on Mathematical Foundations of Computational Anatomy-Geometrical and Statistical Methods for Modelling Biological Shape Variability},
  pages={75--86},
  year={2011}
}

@article{rogers1987structure,
  title={The structure of the vorticity field in homogeneous turbulent flows},
  author={Rogers, Michael M and Moin, Parviz},
  journal={Journal of Fluid Mechanics},
  volume={176},
  pages={33--66},
  year={1987},
  publisher={Cambridge University Press}
}

@inproceedings{kurz2015heart,
  title={Heart phase estimation using directional statistics for robotic beating heart surgery},
  author={Kurz, Gerhard and Hanebeck, Uwe D},
  booktitle={2015 18th International Conference on Information Fusion (Fusion)},
  pages={703--710},
  year={2015},
  organization={IEEE}
}

@article{laio2007verification,
  title={Verification tools for probabilistic forecasts of continuous hydrological variables},
  author={Laio, Francesco and Tamea, Stefania},
  journal={Hydrology and Earth System Sciences},
  volume={11},
  number={4},
  pages={1267--1277},
  year={2007},
  publisher={Copernicus GmbH}
}

\appendix
\section{Minimum wrapping number specification} \label{app:kmin}
An identifiability issue stems from joint inference of $K$ and $Z$, as there are infinite pairs of instances of the two for which the likelihood $p(Y\mid Z,K)$ is equivalent.  While this is not an issue for the purposes of inferring the wrapped space $Y$, it does mean the choice of $k_{\min}$, which determines the smallest possible wrapping number, is important for both fitting and prediction.  The most appropriate choice for $k_{\min}$ should depend on the lower range of $Z$, which can be difficult to know prior to fitting since the latent unwrapped GP is assumed to be real-valued.  However, prior specification for the intercept term $\alpha$, which is also used as its starting value, has a strong influence on the expected range of $Z$.

\begin{figure}[ht!]
\centering
\includegraphics[width=0.45\textwidth, trim=0 10 0 50, clip=TRUE]{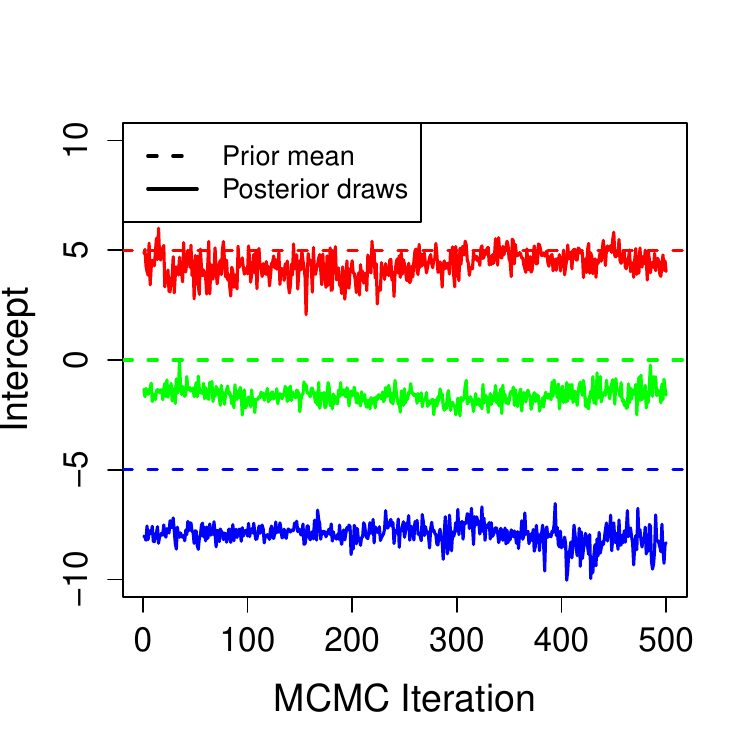}
\includegraphics[width=0.45\textwidth, trim=0 10 0 50, clip=TRUE]{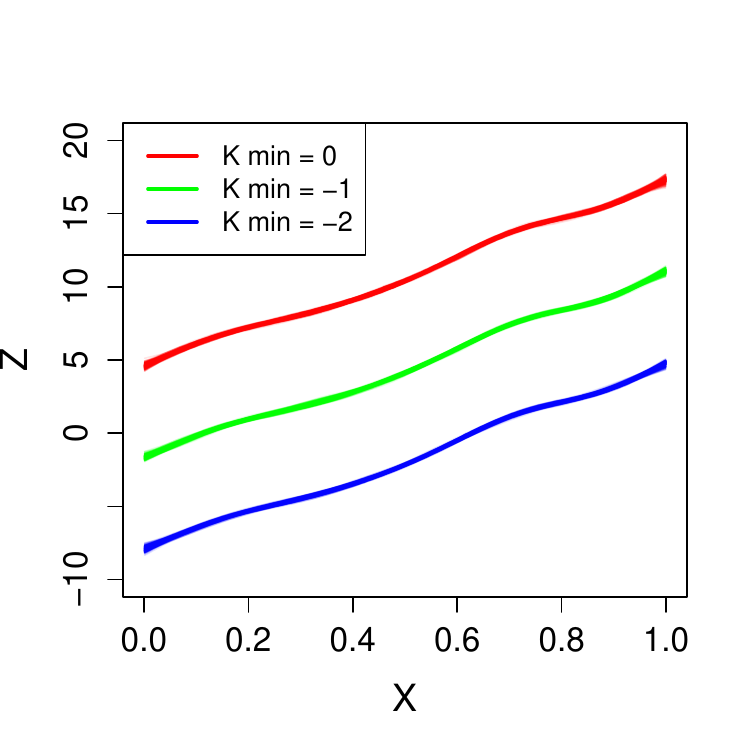}
\caption{{\em Left:} Posterior intercept draws for three separate fits, each with different prior mean specifications shown by the dashed line of the same color. {\em Right:} Posterior $Z$ draws for these three models, each with a different $k_{\text{min}}$.}
\label{fig:kmin}
\end{figure}
As an illustrative example, consider three different prior mean specifications for $\alpha$ shown in the left panel of Figure \ref{fig:kmin}.  The corresponding posterior draws for $\alpha$ after being trained on the same data from Figure \ref{fig:ex_fit} settle on three different regions of the space, each of which is $2\pi$ apart.  The right panel shows the posterior draws of the latent space for these three fits, each of which with a different value for $k_{\min}$.

While prior mean specification of $\alpha$ gives insight into which $k_{\min}$ is appropriate, it can be challenging to infer where exactly $Z$ will ``settle'' in the MCMC.  We therefore allow for an appropriate change in $k_{\min}$ at an iteration that is typically removed as a burn-in.  Setting $k_{\min} = \min\{z_1^{(t)},\dots,z_n^{(t)}\} \text{ mod } 2\pi$ for sufficiently large $t$ enables $K^{(t)}$ to properly shift up or down based on $Z^{(t)}$.  We have found that specifying a prior for $\alpha$ centered around zero, initializing $k_{\min} = -2$, and resetting $k_{\min}$ at $t=1{,}000$ provides consistent appropriate estimation of $Z$ and $K$ for all of our examples in our work.

\end{document}